\documentclass[10pt,a4paper]{article}
\usepackage[utf8]{inputenc}
\usepackage{jcappub}
\usepackage{amsmath}
\usepackage{amsfonts}
\usepackage{amssymb}
\usepackage{calc}
\usepackage{mathtools}
\usepackage{mathabx}
\usepackage{cancel}
\usepackage[dvipsnames]{xcolor}
\usepackage{comment}
\usepackage{graphicx}
\graphicspath{{images/}}
\usepackage{pgfplotstable}
\usepackage{soul}
\usepackage{slashed} 

\newcommand{\bell}{\boldsymbol{\ell}}

\newcommand{\be}{\begin{equation}}
\newcommand{\ee}{\end{equation}}
\newcommand{\beq}{\begin{equation}}
\newcommand{\eeq}{\end{equation}}
\newcommand{\bea}{\begin{eqnarray}}
\newcommand{\eea}{\end{eqnarray}}

\newcommand{\bx}{{\mathbf  x}}

\newcommand{\bL}{{\mathbf  L}}
\newcommand{\bn}{{\mathbf  n}}

\newcommand{\bfe}{{\mathbf  e}}

\newcommand{\bV}{{\mathbf V}}

\renewcommand{\bell}{\boldsymbol{\ell}}

\newcommand{\bnabla}{{\boldsymbol{\nabla}}}

\newcommand{\dd}{\partial}
\newcommand{\dr}{{\rm d}}
\newcommand{\HH}{{\cal H}}

\newcommand{\lan}{{\langle}}
\newcommand{\ran}{{\rangle}}

\newcommand{\OO}{{\cal O}}

\newcommand{\ga}{\gamma}

\newcommand{\de}{\delta}
\newcommand{\De}{\Delta}

\newcommand{\Om}{\Omega}

\newcommand{\cd}{\cdot}
\newcommand{\ra}{\rightarrow}
\newcommand{\pa}{\parallel}

\definecolor{magenta}{rgb}{0.1,0.98,0.6}

\definecolor{dgreen}{rgb}{0,0.6,0.0}

\definecolor{dblue}{rgb}{0,0.0,0.7}
\definecolor{dpurple}{rgb}{0.5,0.0,0.5}

\allowdisplaybreaks

\title{ An Estimator for the lensing potential from galaxy number counts}
\author[a]{Viraj Nistane}
\author[b,c,d]{Mona Jalilvand }
\author[a]{Julien Carron}
\author[a]{Ruth Durrer}
\author[a]{Martin Kunz}

\affiliation[a]{
Universit\'e de Gen\`eve, D\'epartement de Physique Th\'eorique and Centre for Astroparticle Physics,
24 quai Ernest-Ansermet, CH-1211 Gen\`eve 4, Switzerland}
\affiliation[b]{ Department of Physics, McGill University, 3600 rue University, Montreal, QC H3A 2T8, Canada}
\affiliation[c]{ McGill Space Institute, McGill University, 3550 rue University, Montreal, QC H3A 2A7, Canada}
\affiliation[d]{Institute of Theoretical Astrophysics, University of Oslo, 0315 Oslo, Norway}
\emailAdd{viraj.nistane@unige.ch}
\emailAdd{mona.jalilvand@mail.mcgill.ca}
\emailAdd{ruth.durrer@unige.ch}
\emailAdd{martin.kunz@unige.ch}

\date{\today}

\abstract{We derive an estimator for the lensing potential from galaxy number counts which contains a linear and a quadratic term. We show that this estimator has a much larger signal-to-noise ratio than the corresponding estimator from intensity mapping. We show that this is due to the additional lensing term in the number count angular power spectrum which is present already at linear order. We estimate the signal-to-noise ratio  for future photometric surveys. We find that particularly at high redshifts, $z\gtrsim 1.5$, the signal to noise ratio can become of order 30. We therefore claim that number counts in photometric surveys are an excellent means to measure tomographic lensing spectra.}

\begin{document}

\maketitle

\section{Introduction}
Light coming to us from far away sources is deflected by the intervening gravitational field due to cosmic structure. In the regime of weak lensing, to first order in the cosmological perturbations, this can be described by the lensing potential
\be
\phi(\bn,z) = -\int_0^{r(z)} \hspace{-2mm}\dr r\frac{r(z)-r}{r(z)r}\left[\Phi(-r\bn,t_0-r)+\Psi(-r\bn,t_0-r)\right]  \,.
\ee
Here $\Phi$ and $\Psi$ are the Bardeen potentials, $r(z)$ is the comoving distance out to redshift $z$ and $t=t_0-r$ is conformal time along the light path. We neglect possible contributions from tensor perturbations, i.e.\ gravitational waves, as well as from vector perturbations since they are generally small~\cite{Yamauchi:2012bc,Adamek:2015mna}.  For non-relativistic matter and a cosmological constant  the two Bardeen potentials are equal and correspond to the Newtonian gravitational potential.
Light from a source at redshift $z$, seen in {direction $-\bn$} is coming to us from the  angular position $\bn+\bnabla\phi(\bn,z)$, where $\bnabla$  denotes the 2D gradient on the unit sphere and $\bnabla\phi(\bn,z)$ is the deflection angle.

The shear $\ga$,  given by the traceless part of the second derivatives of $\phi$, can be measured via weak lensing of galaxy shapes~\cite{Bartelmann:1999yn}, see~\cite{DES:2021wwk,DES:2020daw,KiDS:2020suj} for recent observational results. 
However, shear measurements are plagued by intrinsic alignment as a serious systematic effect~\cite{Hirata:2004gc,Kirk:2010zk}, and it would be very useful to have a second, alternative measurement of the lensing potential at different redshifts. So far, it has been shown that galaxy surveys can be used to measure the correlation function $\lan \phi(\bn,z)\delta(\bn',z')\ran$ where $\de$ denotes the density contrast (relative matter overdensity) and $z'<z$~\cite{Montanari:2015rga}. 

An alternative approach is to apply \textit{quadratic estimators} to observations of the cosmic microwave background (CMB) radiation \cite{Hu:2001kj,Okamoto:2003zw,Lewis:2006fu}. This method has been very successfully used to reconstruct maps of the lensing potential out to the CMB redshift, i.e., $\phi(\bn,z_{\rm dec})$, $z_{\rm dec} \simeq 1060$, see~\cite{Planck:2013mth,Planck:2018lbu}. 
From the distribution of galaxies, it should be possible to similarly measure $\phi(\bn,z)$ for many different redshifts from $z\sim 0$ up to $z\sim 5$, or even higher using intensity mapping (IM), which is affected by lensing at second order, like the CMB~\cite{Foreman:2018gnv}.

In this paper we extend the idea of a quadratic estimator to galaxy number counts.
We derive an estimator for galaxy number counts which contains also a linear term, as number counts are affected by lensing already at linear order. Furthermore, the quadratic part contains an additional term which, as we shall see, is positive definite and therefore typically larger than the quadratic term which is also present in intensity mapping.  We compare our estimator with the one for intensity mapping and discuss its applicability to planned galaxy surveys.
We obtain promising results for the expected signal-to-noise ratio (SNR) for the next generation of large photometric galaxy surveys. We consider specifically two photometric survey scenarios: (i) a 15'000 square-degree survey with a limiting magnitudes of 27 and 25, based on the specifications for the Legacy Survey of Space and Time that is planned for the Vera C. Rubin observatory \cite{Abell:2009aa} and that we will denote as `LSST-like', and (ii) a 15'000 square-degree photometric survey with limiting depth of 24 modeled on the ESA Euclid satellite mission \cite{Laureijs:2011gra}, that we will call `Euclid-like' in the following sections. For these two scenarios we predict a total SNR of about 38 for each of the surveys considered.
Therefore, near-future galaxy number count observations will provide an excellent means to measure the lensing potential tomographically, in a way that is complementary  to cosmic shear surveys. 

In the next section we derive the  estimator for galaxy number counts. We then compare our result with the one for intensity mapping in Section~\ref{s:int}. In Section~\ref{s:s/n} we estimate the signal to noise for the above-mentioned experimental situations and in Section~\ref{s:con} we conclude. Some detailed derivations are deferred to appendices.

\section{An estimator for the lensing potential from galaxy number counts}\label{s:quad}

Let us first introduce the general philosophy of an estimator: We consider a stochastic observable $X$ at redshift $z$ which is affected at first and second order by the lensing potential. We consider terms linear in the lensing potential multiplied or not with the unlensed signal which is denoted by $\tilde X$. We assume that terms quadratic in the lensing potential can be neglected. We  work in the flat sky approximation which is sufficiently accurate if we assume that our survey has small angular extent and can be considered at a fixed direction far away at roughly fixed redshift $z$. A point $r(z)\bn$ on our survey can then be denoted by $r(z)\bn \simeq r(z)(\bfe + \bx)$ where $\bx$ is a small (dimensionless) vector normal to the mean direction $\bfe$ of our survey. At the end we shall integrate over the finite thickness of our redshift bin. 

We denote the (unitary) Fourier transform of our variable by
$$
\tilde X(\bell,z) = \int \frac{\dr^2x}{2\pi} \tilde X(\bx,z)\exp(i\bx\cd\bell) \,.
$$
Statistical isotropy and homogeneity  imply\footnote{We use CLASS \cite{Blas:2011rf,DiDio:2013bqa} (\url{http://class-code.net/}) to compute spectra except where we say otherwise.}
\bea\label{e:Xiso}
\lan\tilde X(\bell,z)\tilde X(\bell',z')\ran&=& \de(\bell+\bell') \tilde C_\ell(z,z') \,.
\eea
Here $\tilde X$ is the {\em unlensed} variable. In this section we  mainly consider equal redshifts, $z=z'$, but the generalization to unequal redshifts (cross-correlations) is straight forward.

In $\ell$ space weak lensing generically affects $X$ through
a linear term and a convolution with some kernel $K_X$ which depends on the variable $X$ we consider, such that
\be\label{e:Xlens}
X(\bell,z)= \tilde X(\bell,z) + g_X(\ell,z)\phi(\bell,z) +  \int \frac{d^2 \ell'}{2\pi}K_X(\bell',\bell,z) \tilde X(\bell',z) \phi(\bell-\bell',z) + \OO(\phi^2)  \,.
\ee
The convolution in \eqref{e:Xlens} is simply the generic form of a product in real space, combined with derivatives which determine the form of $K_X$. 
We also assume that $K_X$ is parity symmetric, $K_X(\bell',\bell)=K_X(-\bell',-\bell)$. There will in general also be higher order terms (in $\phi$) which we neglect in our approach. We assume lensing to be weak and to have a small impact on our variable $X$. The second term, linear in $\phi$, is not present in the CMB and intensity mapping. Here $g_X(\ell,z)$ is some generic, deterministic function of the Fourier mode $\ell$ and redshift $z$, we shall specify it for galaxy number counts below. Eq.~\eqref{e:Xlens} implies that for fixed $\phi$, the expectation value of $X$ no longer vanishes but is offset by the presence of $\phi$.

We now introduce  the function $f_X$ as
\be\label{e:def-fX}
f_X(\bell_1,\bell_2,z)=K_X(-\bell_1,\bell_2,z)\tilde C_{\ell_1}(z)+K_X(-\bell_2,\bell_1,z)\tilde C_{\ell_2}(z)
\ee
By definition $f_X(\bell_1,\bell_2,z) = f_X(\bell_2,\bell_1,z)$. 
We define the expectation value $\lan\cdots\ran_\phi$ as an ensemble average only over $X$, at fixed lensing potential $\phi$. This makes sense only if $\phi$ is (nearly) uncorrelated with the stochastic variable $X$.
For sufficiently high redshifts this is usually a good approximation as the lensing kernel peaks roughly in the middle between $0$ and $r(z)$.
It is straightforward  to verify that, considering $\phi$ fixed and taking an expectation value over $X$, neglecting terms quadratic in $\phi$, we obtain (for $\bL\neq 0$)
\bea\label{e:XX1}
\langle X(\bL)\rangle_\phi &=& g_X(L,z)\phi(\bL) \, ,\\ 
\langle X(\bell)X(\bL-\bell)\rangle_\phi &=& \frac{1}{2\pi} f_X(\bell,\bL-\bell)\phi(\bL) \,.  \label{e:XX2}
\eea
As the expectation value of $\tilde X$ vanishes, the offset $g_X(\bell,z)\phi(\bell,z)$ does not contribute to \eqref{e:XX2}.
We can now derive an  estimator for $\phi(\bL)$ which combines the linear and the quadratic terms in $X$ to which $\phi$ contributes. It is given by
\bea
\hat\phi_X(\bL,z) &=& A_X(L,z)N_X(L,z)\int  \frac{d^2\ell}{2\pi}X(\bell,z)X(\bL-\bell,z) F_X(\bell,\bL-\bell,z) \nonumber \\
&& +(1-A_X(L,z))\frac{X(\bL,z)}{g_X(L,z)} \label{e:estX}\\
\mbox{with}&& \nonumber \\
F_X(\bell_1,\bell_2,z) &=& \frac{f_X(\bell_1,\bell_2,z)}{2C_{\ell_1}(z)C_{\ell_2}(z)}  \label{e:FX} \, , \\
N_X(L,z) &=&\left[\int \frac{d^2\ell}{(2\pi)^2} f_X(\bell,\bL-\bell,z)F_X(\bell,\bL-\bell,z)\right]^{-1}\, ,\label{e:noiseX} \\
A_X(L,z) &=& \frac{C_L(z)}{g_X(L,z)^2N_X(L,z) +C_L(z)}\, .
\eea
By construction $\lan\hat\phi_X(\bL,z)\ran_\phi =\phi(\bL,z)$.
Here we choose $F_X$ and $N_X$ such that the quadratic part of the estimator is unbiased and has minimum variance and we can see that $N_X$ is the noise of the quadratic term.  Note that when ensemble averaging also over $\phi$, we of course obtain $\langle\hat\phi_X(\bL)\rangle =0$.
Also, for reasons of statistical isotropy, $F_X(\bell,\bL-\bell)$ and $f_X(\bell,\bL-\bell)$ depend on directions only via $\bell\cdot\bL$. Therefore, $N_X$ does not depend on the direction of $\bL$. The factor  $A_X$ is chosen to minimize the variance of $\hat\phi$, see Appendix~\ref{A:basic} for details. We have assumed that the $\phi$ power spectrum, which is quadratic in $\phi$, is smaller than both, $C_L$ and $N_X$ and can be neglected in these expressions. We see, not surprisingly, that if $N_X$ is large, $A_X$ is small and more weight is given to the linear term, while if $C_L/g_X^2$ is much larger than $N_X$, the noise of the linear term dominates and $A_X(L,z)$ is close to $1$.
Note that while the $\tilde C_\ell$'s appearing in $f_X$ are the theoretical spectra neglecting lensing, those appearing in $F_X$ are the measured $C_\ell$'s, including both, lensing and noise. The total noise from the combined linear and quadratic terms then becomes
\be
N_X^{\rm (tot)}(L,z)  = \frac{C_L(z)N_X(L,z)}{C_L(z)+g_X^2(L,z)N_X(L,z)}\,. \label{e:NtotX}
\ee
A self-contained derivation of this expression is presented in detail in Appendix~\ref{A:basic}; it can also be obtained directly using linear response on the likelihood function for a modulated Gaussian signal as discussed in Appendix~\ref{A:like}. There we compare this noise with the one for a toy simulation generating a galaxy catalog by Poisson sampling a Gaussian galaxy density field generated from spectra calculated with CAMB\footnote{\url{http://camb.info}}~\cite{Lewis:1999bs}. The agreement is excellent.

In the limit $g_X\ra 0$ we recover the quadratic noise, $N_X$, while in the limit of very large $g_X$ the noise tends to $ N_X^{\rm (tot)}(L) \ra C_L/g^2_X(L)$. 
\vspace{0.2cm}

Up to the new linear term which is not present in previous derivations, this procedure has been successfully applied to the CMB temperature fluctuations and polarisation~\cite{Planck:2015aco,Planck:2018lbu} and has recently also been proposed for intensity mapping~\cite{Foreman:2018gnv}. In this work we extend it to galaxy number counts. We shall find that the first order term which leads to a combination of linear and quadratic estimator is very important and significantly improves the signal to noise ratio. 

We now apply this formalism to galaxy number counts. Neglecting large scale relativistic effects which are relevant only at very large scales, the number counts at first order in perturbation theory are given by~\cite{Bonvin:2011bg,Challinor:2011bk}
\be
\label{galaxy1stOrder}
\Delta_{g}(z,\mathbf{n})= b_g(z) \delta -\HH^{-1}\bn\bnabla(\bn\cdot\bV)- \left(2-5s(z)\right) \kappa(z, \mathbf{n}) =\tilde\Delta_{g}(z,\mathbf{n})- \left(2-5s(z)\right) \kappa(z, \mathbf{n})\,.
\ee
The first two terms are the density fluctuation and the redshift space distortion (RSD) which we collect as $\tilde \Delta_{g}$ or $\Delta^{\rm std}_{g}$ as they are also called the `standard terms'. The third term is proportional to the convergence,
\be
\kappa(z, \mathbf{n}) =-\frac{1}{2}\De_2\phi(z, \mathbf{n}) \,,
\ee
where $\De_2$ denotes the 2D Laplacian on the sphere.
The term $2$ in the pre-factor $(2-5s(z))$ of convergence in \eqref{galaxy1stOrder} takes into account the convergence of light rays due to lensing which lowers the number of galaxies per apparent surface area while the term $5s(z)$ accounts for the increase due to the enhancement of the flux in a flux limited sample. Here $s$ is the logarithmic derivative of the number density at the flux limit, $F_*$, of the survey, which corresponds to the luminosity $L_*(z)=4\pi D_L(z)^2F_*$ where $D_L(z)$ denotes the luminosity distance,
\be\label{e:szF}
5s(z,F_*) = 2\left.\frac{\dd\log \bar n(z,L)}{\dd\log L}\right|_{L=L_*(z)} \,.
\ee

To obtain the observed $C_\ell$ we have to add noise to the theoretical signal. Apart from cosmic variance, shot noise is usually the dominant noise for number count fluctuations, and we shall include only these noise terms in the present analysis. For a redshift bin with a total number of $n(z)$ galaxies in the bin and a sky coverage $f_{\rm sky}$ the shot noise power spectrum is the inverse of the angular number density,
\be
C_\ell^{\rm SN}  = \left(\frac{n(z)}{4\pi f_{\rm sky}}\right)^{-1}\,.
\ee

To first order in perturbation theory, the lensed, observed $C_\ell$'s are then given by
\begin{align}
C_\ell(z) = \Tilde{C_{\ell} }\left(z\right)+\frac{1}{4}\left(2-5 s\left(z\right)\right)^{2}(\ell(\ell+1))^{2} C_{\ell}^{\phi \phi}\left(z\right) 
 -\left(2-5 s\left(z\right)\right) \ell(\ell+1) C_{\ell}^{{\rm std}\, \phi}\left(z\right) + C_\ell^ {\rm SN} \,.
\end{align}
As before, the superscript `std' indicates the standard contributions from density and redshift space distortions. We include RSD even for relatively wide redshift bins as some of us have shown in \cite{Jalilvand:2019brk} that they can remain important, particularly at low $\ell$.

While the first order expression \eqref{galaxy1stOrder} is sufficient to compute the variance of the estimator, we want to consider number counts up to second order in perturbation theory for the signal.
At second order we obtain, see~\cite{Bertacca:2014wga,Yoo:2014sfa,DiDio:2014lka,Nielsen:2016ldx}, 
\bea \label{galaxy2ndOrder} 
\Delta_{g}(z,\mathbf{n}) &=& 
\tilde\Delta_{g}(z,\mathbf{n}) - \left(2-5s\right) \kappa(z, \mathbf{n})
- \left(2-5s\right) \tilde\Delta_{g}(z,\bn) \kappa(z,\mathbf{n}) \nonumber \\ &&  \qquad
+  \nabla^a \phi(z,\mathbf{n}) \nabla_a \tilde\Delta_{g}(z,\bn) +\OO(2) \,, \qquad
\eea
where $\OO(2)$ denotes higher order terms in the lensing potential which we neglect. 
Here we also see that modes of $\nabla\phi$ normal to the gradient of $\Delta_{g}$ do not contribute to the remapping term (second line) while all modes of $\phi$ contribute to $\kappa$. In $\ell$-space this reflects itself by the fact that the $\kappa$-term generates a positive definite contribution to the kernel while the sign of the remapping term depends on direction. For this reason, the modulation term $\kappa$ can generate a larger signal than the remapping term as we shall see. In addition to the linear term, the second order expression contains the product $\tilde \De_g\kappa$ as well as  the standard lensing term which is also found in the CMB and intensity mapping calculations, see e.g.~\cite{Lewis:2006fu}.
In $\bell$-space in the flat sky approximation this becomes

\bea
\Delta_{g}(\bell,z) &=& \tilde\Delta_{g}(\bell,z)   - \ell^2 \left(1-\frac{5}{2}s(z)\right)\phi(\bell,z) 
\nonumber \\
&& -\int \frac{d^2\ell_1}{2\pi} \tilde\Delta_{g}(\bell_1,z)\phi(\bell-\bell_1,z)\left[ \left(1-\frac{5}{2}s\right)(\bell-\bell_1)^2+\bell_1 \cdot (\bell-\bell_1)\right] \,.
\eea
Hence, for our case of interest, galaxy number counts, the function $g$ and the kernel $K$ are given by
\bea
g_\De(\ell,z) &=& -\ell^2\left(1-\frac{5}{2}s(z)\right) \, , \\
K_\De(\bell_1,\bell_2,z) &=& -\left(1-\frac{5}{2}s(z)\right)(\bell_2-\bell_1)^2 -\bell_1\cdot(\bell_2-\bell_1) \,.
\label{e:KDelta}
\eea
The second term of \eqref{e:KDelta} is the kernel of CMB lensing and also intensity mapping, but the first term is new and only present for number counts. Note that this first term is always negative while the sign of the second term depends on the orientation of $\bell_1$ and $\bell_2$. Also new is of course the entire first order term.

For the ensemble average at fixed lensing potential $\phi$ this yields
\bea
   \lan\Delta_{g}(\bell,z)\ran_\phi &=& g_\De(\ell,z)\phi(\bell,z) \, ,\\
    \lan\Delta_{g}(\bell,z)\Delta_{g}(\bell',z)\ran_\phi &=& \delta(\bell+\bell') \tilde C_\ell(z)  -\frac{1}{2\pi}\phi(\bell+\bell') \,\times \nonumber \\
    && \hspace{-4cm}\left[ \left(1-\frac{5}{2}s\right)(\bell+\bell')^2(\tilde C_{\ell'}(z) +\tilde C_\ell(z) )- \bell'\cdot (\bell+\bell')\tilde C_{\ell'}(z)  - \bell\cdot(\bell'+\bell)\tilde C_\ell(z) \right] 
+\OO(\phi^2) \,, \qquad \label{e:kern1}
\eea
where $\tilde C_\ell(z) $ denotes the power spectra from the standard terms, $\tilde\De_g$, i.e. neglecting lensing convergence.  The first order contribution $\ell^2\left(1-5s/2\right)\phi(\bell,z)$ disappears in the quadratic expectation value  since $\lan \tilde\Delta_{g}(\bell',z)\ran_\phi=0$.

From \eqref{e:kern1} we can read off the number count kernel :
\bea
f_\De(\bell_1,\bell_2,z) &=& -\left(1-5s/2\right)(\bell_1+\bell_2)^2\left(\tilde C_{\ell_1}(z) +\tilde C_{\ell_2}(z)\right ) 
\nonumber \\  && 
+\bell_1\cdot(\bell_1+\bell_2)\tilde C_{\ell_1}(z) + \bell_2\cdot(\bell_1+\bell_2)\tilde C_{\ell_2}(z) \,,  \label{e:kern2}\\
\mbox{and}&&\nonumber\\
F_\De(\bell_1,\bell_2,z) &=& \frac{f_\De(\bell_1,\bell_2,z)}{2C_{\ell_1}(z)C_{\ell_2}(z)} \,. \label{e:FDe}
\eea
Comparing this kernel to the one for intensity mapping~\cite{Foreman:2018gnv} we find that in the number counts have  additional contributions proportional to $(2-5s)$. Most importantly there is the additional linear term, but also the quadratic term has an new contribution with this pre-factor. As already mentioned, this term is absent in intensity mapping since lensing conserves surface brightness, which is equivalent to setting $s=2/5$ for measurements of surface brightness.  The same is true for the CMB temperature anisotropies. Also there, the kernel is given solely by the second term of~\eqref{e:kern2}. 
Since the CMB comes from one definite surface there is no redshift dependence. Only $\phi(\bn,z_{\rm dec})$ can be determined from the CMB. 
Instead, for intensity mapping and for galaxy number counts, good resolution in redshift allows us to measure $\phi(\bn,z)$ in several distinct redshift slices. The noise for a fixed mode $\bL$ at redshift $z$ is given by

\bea
N_\De^{\rm (tot)}(L)  &\simeq & \frac{C_L N_\De(L)}{C_L+L^4\left(1-\frac{5}{2}s(z)\right)^2N_\De(L))}  \nonumber \\  &=& \left(\frac{L^4\left(1-\frac{5}{2}s(z)\right)^2}{C_L}+\frac{1}{N_\De(L)}\right)^{-1}  \label{e:Ntot}\\
 \mbox{with} \nonumber \\
N_\De(L) &=&\left[\int \frac{d^2\ell}{(2\pi)^2} f_\De(\bell,\bL-\bell,z)F_\De(\bell,\bL-\bell,z)\right]^{-1} \,. \label{e:noiseDe}
\eea
In the last expression of Eq.~\eqref{e:Ntot} we have split the noise into its contribution from the linear estimator and from the quadratic estimator, 
\bea
\frac{1}{N^{\rm (tot)}} &=& \frac{1}{N_\De^{\rm (lin)}} + \frac{1}{N_\De^{\rm (quad)}}\,, \qquad \mbox{where}\\ 
N_\De^{\rm (quad)} &\equiv& N_\De \qquad \mbox{and} \\ N_\De^{\rm (lin)} &\equiv& \frac{C_L}{L^4\left(1-\frac{5}{2}s(z)\right)^2}\,.
\eea
Assuming that $N_\De$ and $C_L$ are of similar order, we might expect that, especially at higher $L$, the linear term will dominate the signal as its noise term decays like $L^4$. However, also the integral $N^{-1}_\De(L)$ scales naively like  $L^4/C_L$ and so it is a priori not clear which noise term dominates. We shall see, however, that in our examples the linear noise is always smaller than the quadratic noise and the signal is therefore dominated by the linear term.

\section{Comparison with intensity mapping}\label{s:int}
\begin{figure}[h!]
    \centering
    \includegraphics[width=0.9\textwidth]{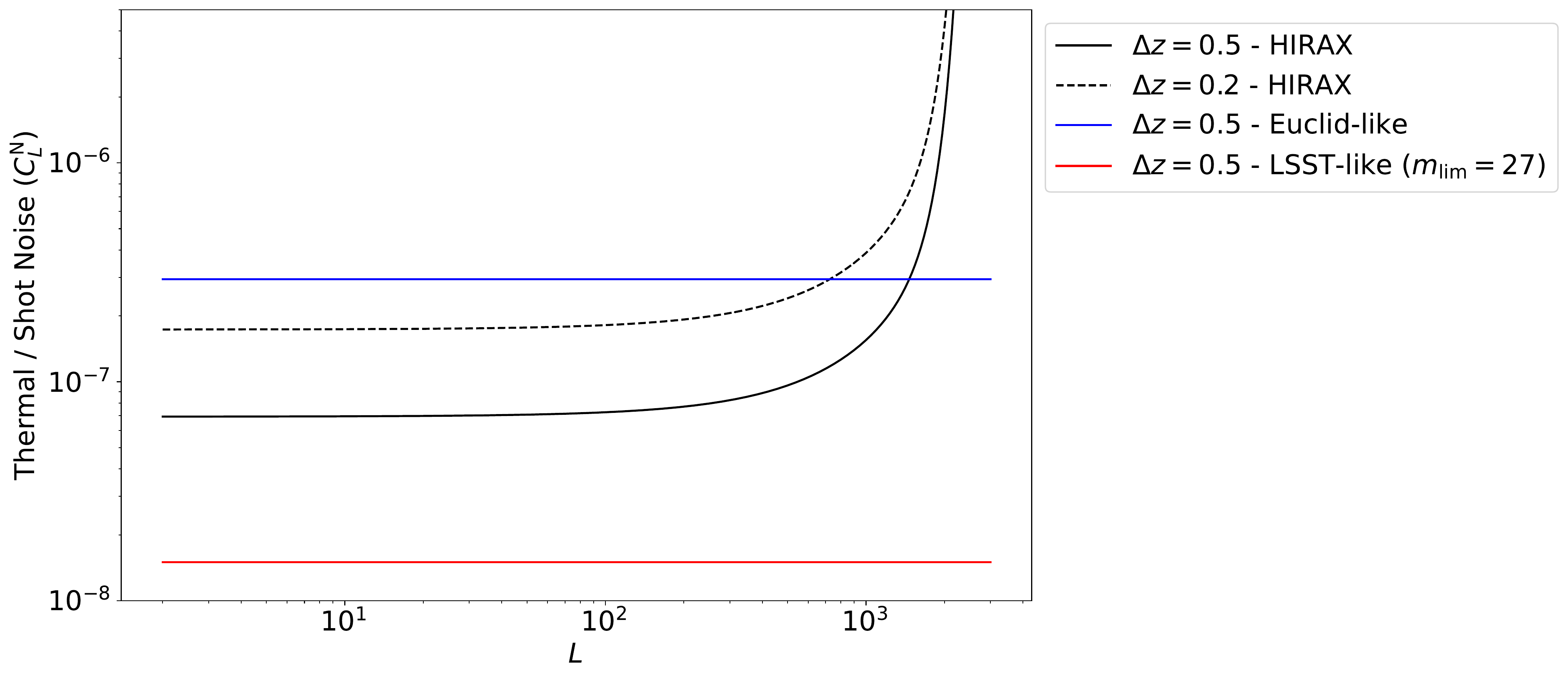}
    \caption{The main noise contributions to IM and number count spectra, thermal noise and shot noise. As an example, we show the predictions at $z=1.9$ for $\Delta z=0.5,0.2$ for the thermal noise of the HIRAX IM survey, and for the shot noise in the Euclid-like photometric and LSST-like ($m_{\lim}=27$) surveys with $\Delta z=0.5$.}
    \label{fig:tn_vs_sn}
\end{figure}

Before calculating the SNR for our estimator, we compare its noise level with the noise from intensity mapping. While shot noise is the dominant noise for number counts, thermal noise is the most relevant noise source for intensity mapping. In Appendix~\ref{C:noise} we briefly discuss thermal noise for intensity mapping and give the relevant ingredients for the Hydrogen Intensity and Real-Time Analysis eXperiment (HIRAX) experiment \cite{Crichton:2021hlc}.
In Fig.~\ref{fig:tn_vs_sn} we compare the thermal noise from an IM survey (HIRAX) with the shot noise from galaxy number counts (LSST-like with $m_{\lim}=27$ and the Euclid-like photometric survey), for a redshift $z=1.91$. The thermal noise from intensity mapping is situated in between the shot noise for these two surveys.
At lower redshifts the number density is typically higher leading to lower noise for the number count signal. But for the examples considered in this paper, we find that thermal noise is always comparable to shot noise.

\begin{figure}[h!]
    \centering
    \includegraphics[width=\textwidth]{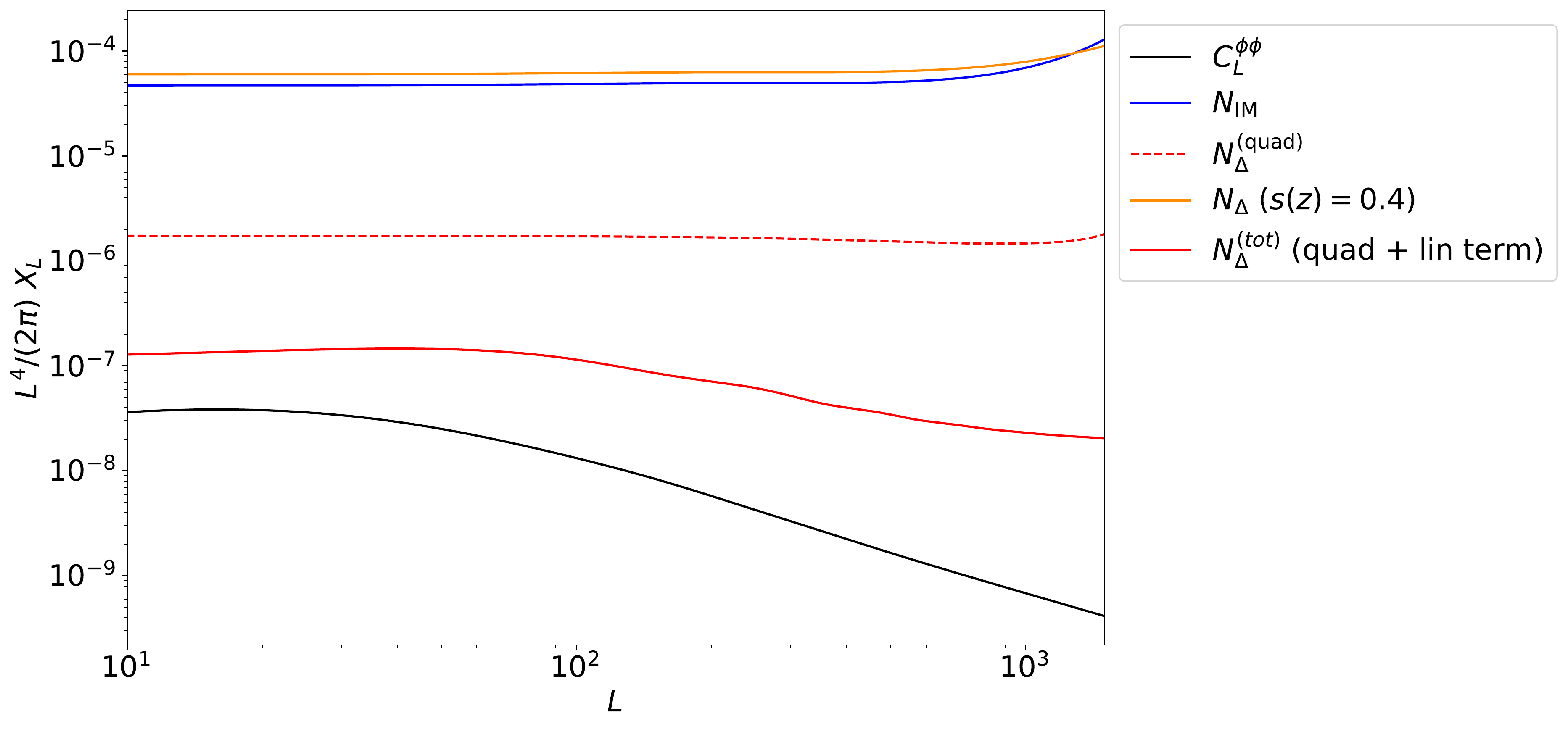}
    \caption{The lensing reconstruction noise for halofit power spectra for IM (HIRAX) and galaxy number counts (Euclid-like) for $z=1.9$ and $\Delta z = 0.5$. We also indicate the signal $C_\ell^{\phi\phi}$ for comparison. All quantities are multiplied by $L^4/(2\pi)$. We also show the galaxy number count noise obtained when replacing $s(z)$ by $2/5$. Note that the naive $L^{-4}$ scaling of the noise holds very well for the quadratic noise, but the total noise, $N_\De^{\rm tot}$ decays faster for $L>60$. This is due to the significantly smaller linear noise.}
    \label{fig:Noise_z_1.91_sw_0.25_linear_galaxy_IM}
\end{figure}
In photometric galaxy surveys we have limited redshift resolution, so that we simply average over a redshift bin. This is an important difference compared to intensity mapping analyses that have an excellent redshift resolution and where one can consider multiple radial modes $k_\pa$ inside a redshift bin. However, IM surveys have to remove the lowest five or so Fourier modes since these are dominated by unresolved foregrounds \cite{Foreman:2018gnv}. The low Fourier modes actually contain the largest contributions to the signal and having to remove them reduces significantly the signal-to-noise ratio  for IM, see \cite{Foreman:2018gnv}. When we compare the two surveys here, we will simply consider the signal from the Fourier zero-mode for both. This over-estimates the SNR for intensity mapping somewhat, but since this paper focuses on galaxy number counts, this is not very relevant here. We will see that number counts have a significantly higher SNR than intensity mapping even when using the Fourier zero mode for both.

To obtain the total noise of the experiment given in Eq.~\eqref{e:Ntot}, we compute the $C_\ell$'s and the function $f_\De$. For the number counts, $f_\De$ is given in Eq.~\eqref{e:kern2}. For intensity mapping, the first term is absent, which corresponds to setting $s=2/5$.

Interestingly, even though the thermal noise is significantly smaller than the shot noise, the total noise level for intensity mapping ($N_{\rm{IM}}$, blue solid line) is more than a factor 10 higher than the noise of the galaxy number counts already for the quadratic term alone ($N_\De^{\rm (quad)}$, red dashed).  The reason for this is twofold. Firstly, $f_\De$ from number counts has an additional term typically of the same order or larger, depending on the value of $s$. For $z>1.5$ we have $s\simeq 1$ so that the pre-factor of this term is $3/2$. Secondly, in the integral \eqref{e:noiseDe}, the $\ell$-factor of the first term in $f_\De$ is simply $L^2$, which is always positive, while the other terms have a pre-factor $\bell\cd\bL$ that has vanishing mean when integrated over $\bell$. Both these facts increase the integral in \eqref{e:noiseDe} and decrease the noise $N_\De$ for galaxy number counts. In Fig.~\ref{fig:Noise_z_1.91_sw_0.25_linear_galaxy_IM} we also indicate the galaxy number count noise for $s=2/5$ where the first term in $F_\De$ vanishes (orange solid line). In this case, the noise is similar to the one for intensity mapping which proves that the better performance of number counts is really due to the additional term in $f_\De$ proportional to $(5s-2)$. Most importantly, however, the noise of the linear term is another factor of more than 10 smaller than the quadratic noise, reducing overall the total noise of the number count estimator (red solid line) by more than two orders of magnitude when compared to intensity mapping.
Despite this significantly reduced noise for galaxy number counts, for each individual mode the noise is still typically more than one order of magnitude larger than the signal, the black solid line in Fig~\ref{fig:Noise_z_1.91_sw_0.25_linear_galaxy_IM}.

Also for other redshifts the quadratic noise of intensity mapping, $N_{\rm{IM}}$, is typically one to two orders of magnitude larger than the one for galaxy number counts. An exception is redshifts where for the galaxy survey $s\approx 2/5$. In this case, the additional term in  $f_\De$ nearly vanishes and the  noise becomes as large as the one of intensity mapping. 
Furthermore, also the linear contribution vanishes in the limit $s\ra 2/5$.
However, considering the definition of $s(z)$, see Eq.~(\ref{e:szF}), one finds that by 
 choosing a somewhat higher flux limit, $F_*$, one can change $s(z)$ and move it away from $2/5$. A drawback of this procedure is that, by increasing the flux limit, we are losing galaxies which increases the shot noise. For each given survey and in each redshift bin there will therefore be a `sweet spot' in $F_{*}$ for which the noise in our estimator is minimal.

\section{Signal to noise for the lensing estimator from galaxy number counts}\label{s:s/n}

In this section we discuss the application of the estimator to two exemplary photometric surveys based on the specifications for the Legacy Survey of Space and Time (LSST) of the Vera C. Rubin Observatory~\cite{Abell:2009aa,Abate:2012za} and for Euclid~\cite{Laureijs:2011gra,Amendola:2016saw}, the ESA satellite under preparation for launch in early 2023.

To evaluate our estimator we need forecasts for the number densities $n(z)$, the galaxy bias, $b(z)$ and the magnification bias $s(z)$. For LSST-like survey, we use the approximations given in Ref.~\cite{Alonso:2015uua}, see also~\cite{Jelic-Cizmek:2020pkh}, while for Euclid-like survey we follow the forecasts of Ref.~\cite{Euclid:2021rez}.
\begin{figure}[h!]
    \centering
    \includegraphics[width=0.7\textwidth]{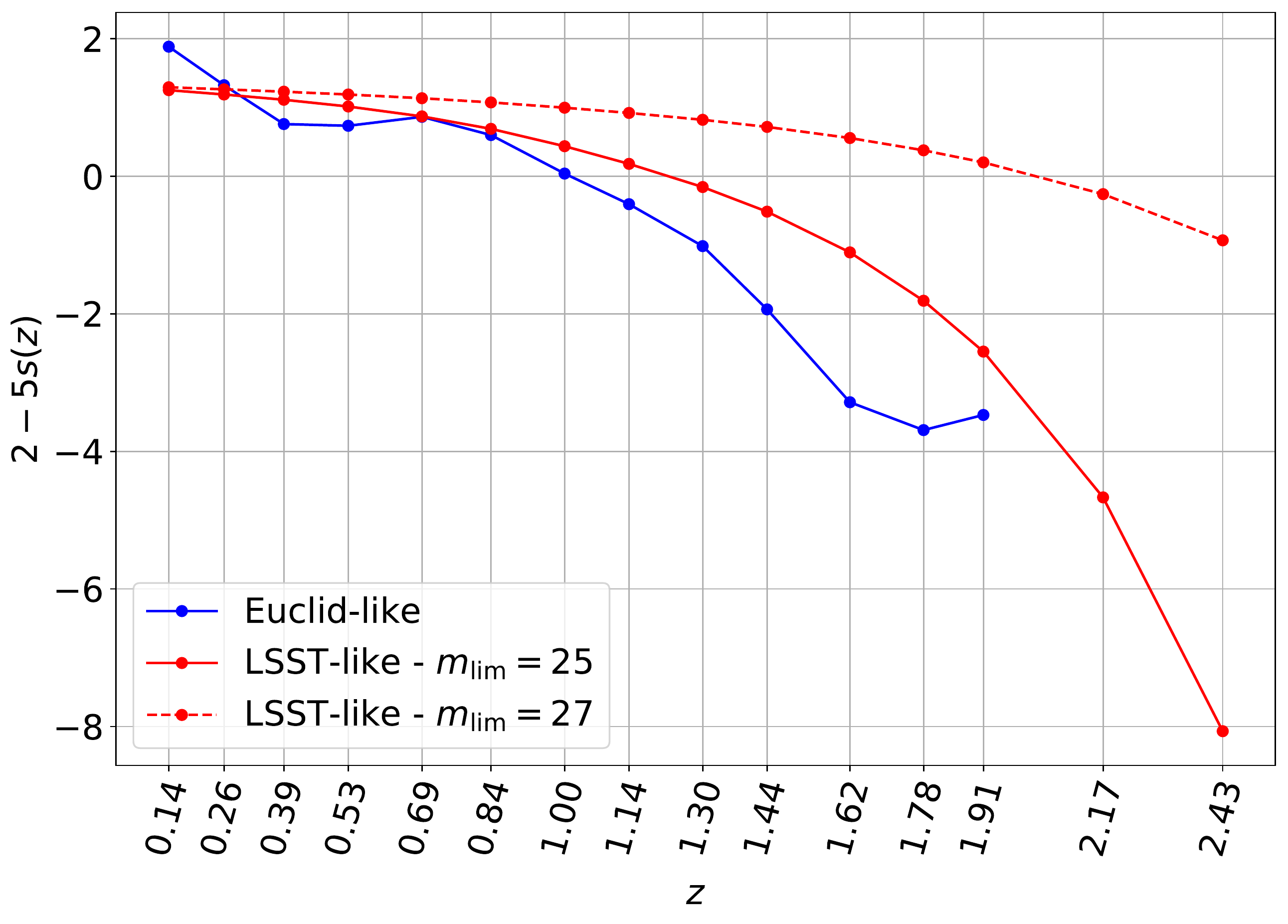}
    \caption{Forecasts for the magnification bias s(z), for the surveys considered in this work, see Refs.~\cite{Alonso:2015uua,Euclid:2021rez}.}
    \label{fig:sz_all}
\end{figure}
In Fig.~\ref{fig:sz_all} we present the pre-factor $2-5s(z)$ as a function of redshift for our two examples of photometric surveys; for LSST-like we consider the two magnitude limits $m_{\lim}= 25$ and $m_{\lim}= 27$. For Euclid-like, the pre-factor passes through $0$ at $z\simeq 1$ while for LSST-like with $m_{\lim}= 25$ this happens at $z\simeq 1.1$.
For LSST-like with $m_{\lim}= 27$ the pre-factor remains positive for $z\lesssim 2$ and it is never much larger than 1. Lensing as an integrated quantity becomes more important at higher redshifts, so we expect that an LSST-like survey with $m_{\lim}= 27$ is at a disadvantage as its value of $|2-5s|$ is much smaller at the high redshift end than the ones for the two other examples.

\begin{figure}[h!]
    \centering
    \includegraphics[width=0.7\textwidth]{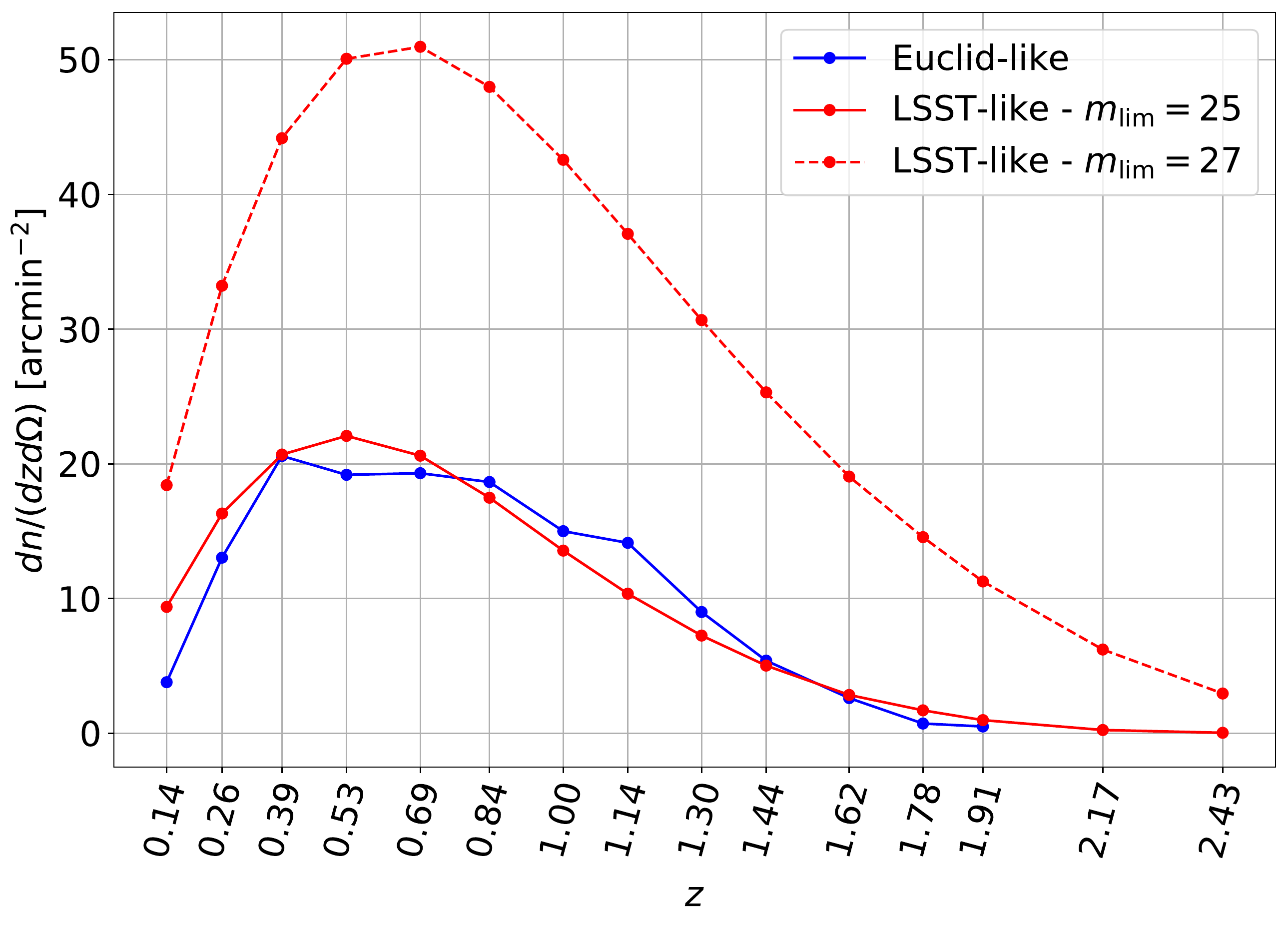}
    \caption{
    {\large{$\frac{dn}{dz d\Omega}$}} [arcmin$^{-2}$] 
    for the  surveys considered in this work, from~\cite{Alonso:2015uua,Euclid:2021rez}.}
    \label{fig:nz_all}
\end{figure}
In Fig.~\ref{fig:nz_all} we plot the predicted galaxy number densities as functions of redshift for the three examples under consideration. 
While both surveys have similar sky fractions, $f_{\rm sky}\simeq 0.35$, LSST can observe a significantly higher number of galaxies, depending on the magnitude limit used. We will however also see that for the values of $\ell$ that we consider here, shot-noise is not very important, so that the higher galaxy density for LSST-like survey with $m_{\lim}= 27$ is not able to overcome the disadvantage from the smaller $|2-5s|$ at high $z$.

To compute the SNR we then calculate the lensed and unlensed $C_\ell$'s which enter our expressions for the noise as well as the lensing power spectrum which is our signal. To determine the quadratic noise we perform the integral \eqref{e:noiseDe} until $L_{\max}=1500$ (which requires calculating the $C_\ell$'s up to 3000).
The SNR for our  estimator for a redshift bin with mean redshift $z$ and a multipole $L$ is then given by
\be
\left(\frac{S}{N}\right)^2(L,z) = 
 \frac{f_{\rm sky}\left(2L+1\right)}{2}\left(\frac{C_L^{\phi\phi}(z)}{C_L^{\phi\phi}(z)+N^{\rm tot}_\De(L,z)}\right)^2 \,.
\ee

\begin{figure}[h!]
    \centering
    \includegraphics[width=0.8\textwidth]{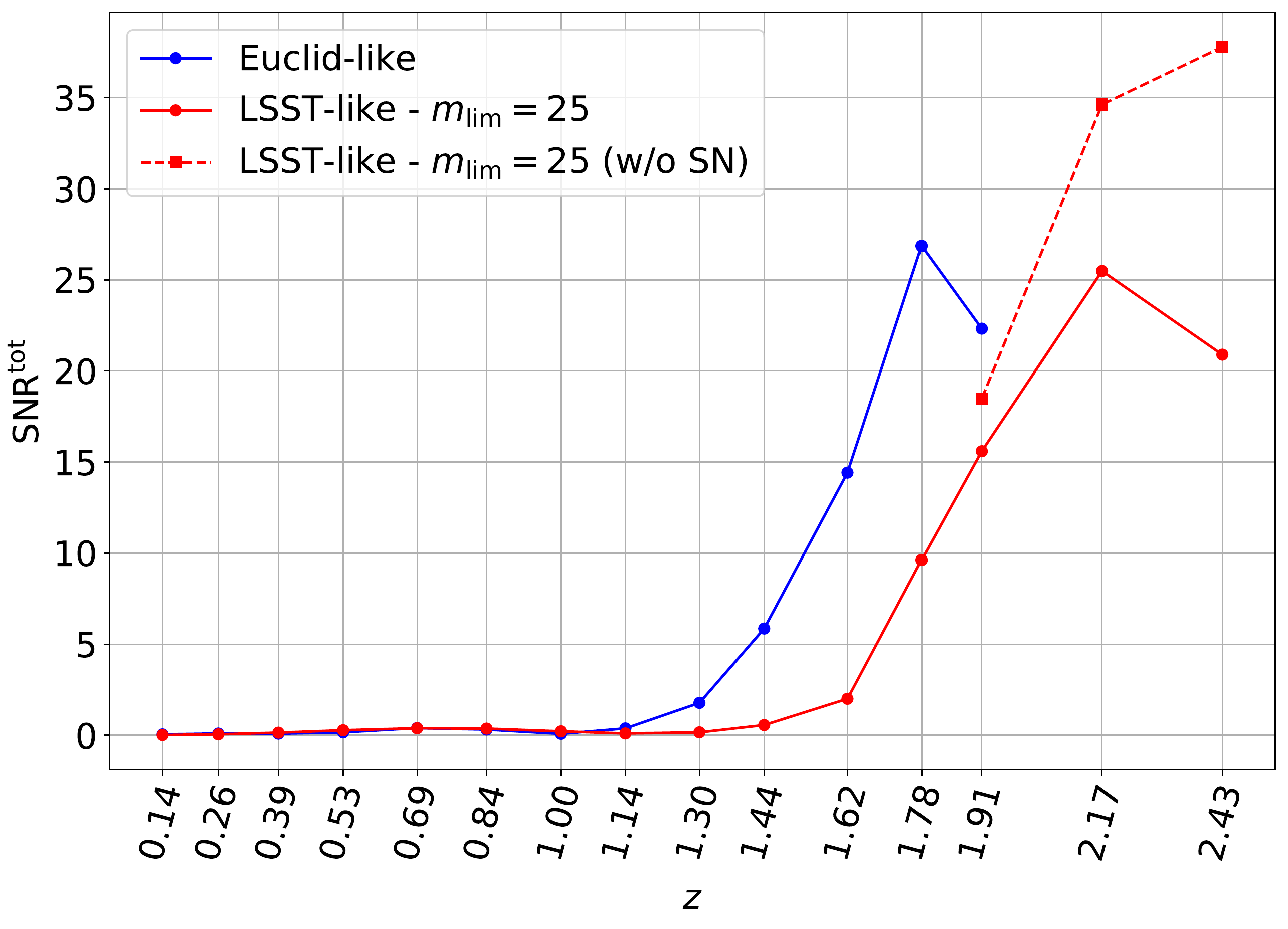}
    \caption{Total SNR per redshift bin plotted against the mean redshift of each bin for non-linear perturbation theory results for Euclid-like and LSST-like ($m_{\lim}=25$) surveys
    representing the $4th$ and $6th$ columns of table~\ref{table:snr_tot}. In the highest redshift bins of the LSST-like survey, shot noise starts to become important, the red-dashed curve shows the SNR that we would find without shot noise.}
    \label{fig:SNRtot_vs_z}
\end{figure}

The first factors estimate the number of independent $M$-modes of the given $L$ in the considered fraction of the sky, $f_{\rm sky}$ while the last factor compares the expected squared signal with the variance of our estimator. The factor $1/2$ is due to the $C_L$'s not being Gaussian but squares of a Gaussian variable, see e.g.~\cite{Durrer:2020fza}. For completeness we have added the cosmic variance, $C_L^{\phi\phi}$, in the noise. Considering a series of values $L$ we can  define the cumulative signal to noise in a given bin with mean redshift $z$ by
\be
\left(\frac{S}{N}\right)_{\rm tot}(z) =\sqrt{\sum_{L=L_{\min}}^{L_{\max}}\left(\frac{S}{N}\right)^2(L,z)} ~ .
\ee
$L_{\min}$ is determined by the sky coverage and should be larger than about 20 as we work in the flat sky approximation, while $L_{\max}$ is determined either by the resolution of our map or by the onset of non-linearities in the number counts which are not included in this treatment.
However, since redshift space distortions are not relevant at higher $\ell$'s in photometric surveys, we can safely use halofit~\cite{Takahashi:2012em} as a good approximation for non-linearities in a photometric survey at least up to $L_{\max}\simeq 1500$ to forecast the SNR (see \cite{Lepori:2021lck} for a comparison of numerical N-body simulations and halofit for photometric number count surveys). We show the result for two cases in Fig.~\ref{fig:SNRtot_vs_z} as a function of redshift and give the numbers for all three examples considered in this work in Table \ref{table:snr_tot}.

\renewcommand{\arraystretch}{1.5}
\begin{table}[!h]
\centering
\begin{tabular}{|c|c||c|c||c|c||c|c|}
\hline
&  &  \multicolumn{6}{c|}{SNR} \\ \cline{3-8}
z & $\Delta$ z  & \multicolumn{2}{c||}{Euclid}   &  \multicolumn{2}{c||}{LSST ($m_{\lim}=25$)}  & \multicolumn{2}{c|}{LSST ($m_{\lim}=27$)} \\ \cline{3-8}
& & linear & halofit & linear & halofit & linear & halofit \\\hline\hline
  0.14 &  0.20 &  0.01 &  0.05 &  0.01 &  0.01 &  0.01 &  0.02\\\hline
  0.26 &  0.20 &  0.04 &  0.10 &  0.03 &  0.05 &  0.04 &  0.06\\\hline
  0.39 &  0.20 &  0.05 &  0.09 &  0.08 &  0.14 &  0.11 &  0.18\\\hline
  0.53 &  0.20 &  0.09 &  0.16 &  0.15 &  0.28 &  0.22 &  0.40\\\hline
  0.69 &  0.20 &  0.21 &  0.40 &  0.21 &  0.39 &  0.36 &  0.69\\\hline
  0.84 &  0.20 &  0.19 &  0.31 &  0.22 &  0.37 &  0.50 &  0.94\\\hline
  1.00 &  0.20 &  0.10 &  0.08 &  0.17 &  0.22 &  0.61 &  1.13\\\hline
  1.14 &  0.20 &  0.21 &  0.38 &  0.12 &  0.10 &  0.66 &  1.19\\\hline
  1.30 &  0.20 &  0.83 &  1.78 &  0.13 &  0.16 &  0.67 &  1.15\\\hline
  1.44 &  0.20 &  2.79 &  5.86 &  0.31 &  0.56 &  0.62 &  1.00\\\hline
  1.62 &  0.20 &  7.41 &  14.42 &  1.05 &  2.01 &  0.50 &  0.71\\\hline
  1.78 &  0.50 &  15.73 &  26.87 &  5.02 &  9.63 &  0.50 &  0.80\\\hline
  1.91 &  0.50 &  14.29 &  22.33 &  8.86 &  15.60 &  0.30 &  0.35\\\hline
  2.17 &  0.50 &  - &  - & 17.66  & 25.49&  0.42  & 0.6 \\\hline
  2.43 &  0.50 &  - &  - & 17.15  & 20.9 &  2.59 & 4.42 \\\hline\hline
\multicolumn {2}{|c||}{$\sqrt{\sum_i{\rm SNR}(z_i)^2}$} & 22.69 &  38.29 &  26.66 &  37.78 &  3.09 &  5.27 \\
\hline
\end{tabular}
\caption{Total SNR for $L_{\max}=1500$ in  different redshift bins for  linear and halofit signals for the Euclid-like and LSST-like surveys. For LSST-like we considered different magnitude limits found in the literature. For this table we set $f_{sky} =0.35$ for both examples. 
We do not include the cross correlations between different bins in this table. Including them in the covariance somewhat reduces the total SNR, e.g from 38.3 to 37.1 for the Euclid-like survey (see Appendix~\ref{A:like}).
\label{table:snr_tot}}
\end{table}
\renewcommand{\arraystretch}{1}

In Table~\ref{table:snr_tot} we show the total SNR inside several redshift bins for $L_{\min}=20$ and $L_{\max}=1500$. At low redshift, on the one hand the lensing signal is low and we therefore expect a relatively low SNR. On the other hand also $s(z)<2/5$ so that the two terms in $f_\De$ have opposite signs. Nevertheless, as long as $s$ is not very close to $2/5$, we still have a significantly larger SNR than from intensity mapping which mainly come from the linear term. The fact that the total SNR is not monotonic with $z$, comes from the pre-factor $2-5s(z)$ which vanishes for $z\sim 1$ in two of our examples and tends to zero for $z\ra 1.9$ in the LSST-like example with $m_{\lim}=27$.
Overall, we find that the most significant bins achieve an SNR of the order of 25, and that the total SNR is about 38 for both Euclid-like and LSST-like ($m_{\rm lim}=25$) surveys individually. As expected, the predicted SNR for LSST-like ($m_{\rm lim}=27$) is much smaller.

\begin{figure}[h!]
    \centering
    \includegraphics[width=0.7\textwidth]{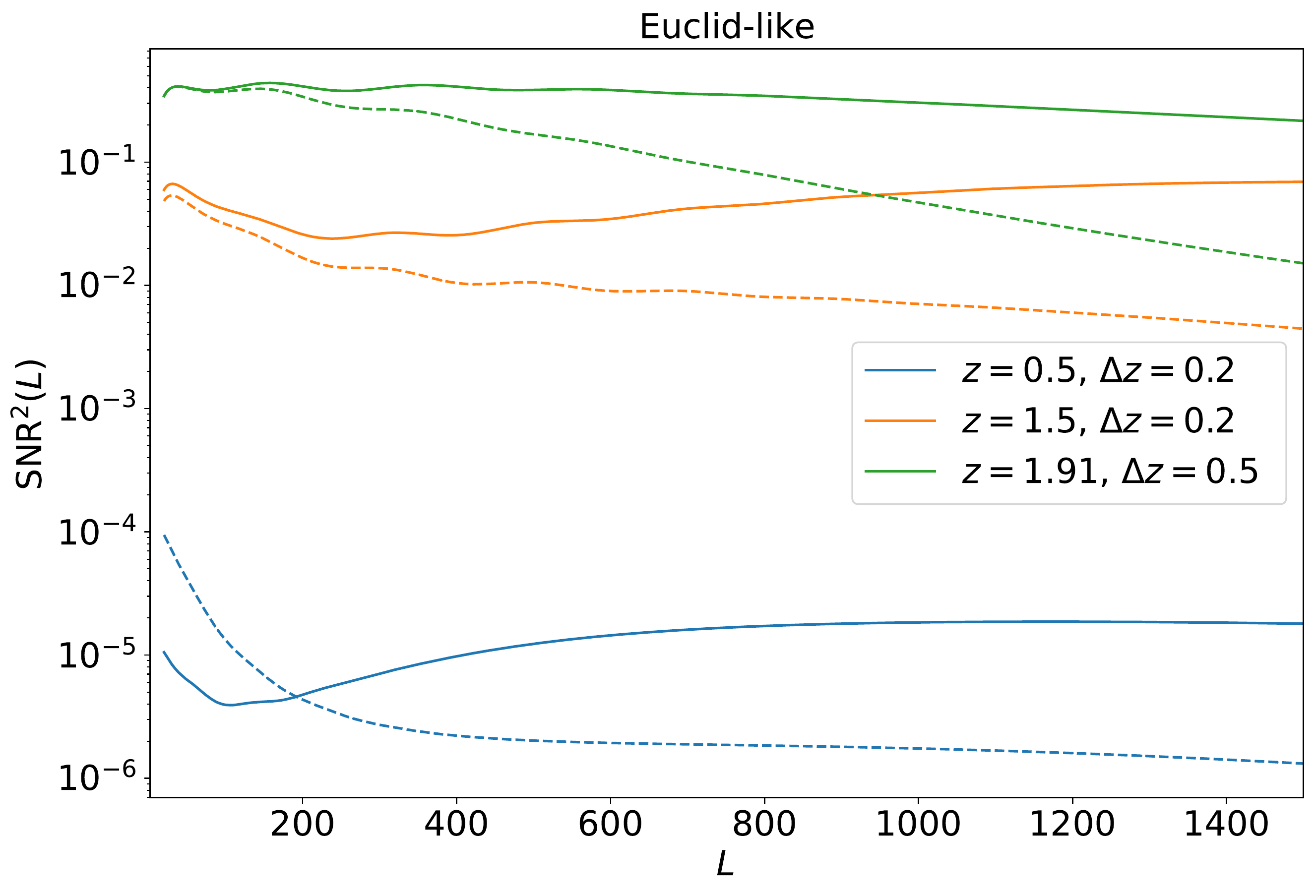}
    \includegraphics[width=0.7\textwidth]{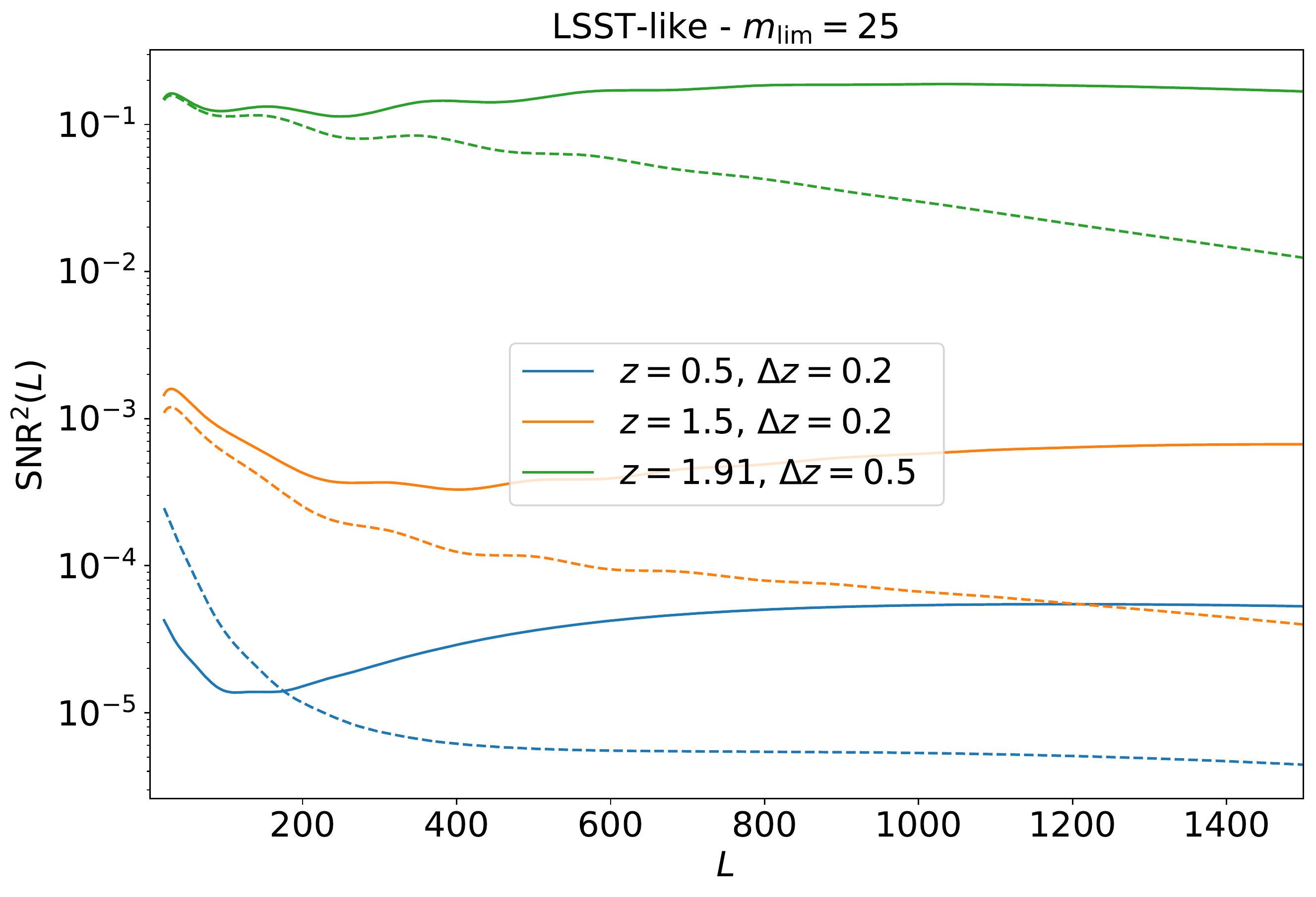}
    \caption{SNR per $L$ for the galaxy number counts. We compare linear perturbation theory results (dotted) with the non-linear results modelled by halofit (solid) for the mean redshifts and bin-widths indicated in the figure.}
    \label{fig:SNR_vs_L_linHaloBoth_galaxy}
\end{figure}

In Fig.~\ref{fig:SNR_vs_L_linHaloBoth_galaxy} we show $(S/N)^2(L,z)$ for three redshift bins as a function of $L$. The signal-to-noise ratio depends on the one hand on the lensing signal which is larger for higher redshifts and for larger values of $|5 s-2|$, and on the other hand on the noise which tends to be larger as well at higher redshifts and on smaller angular scales. Clearly, for both examples the highest redshift bin wins out on all angular scales by one to two orders of magnitude. The low redshift bin $z=0.5$ has a much smaller signal-to-noise. It is also interesting to note that while the SNR from linear perturbation theory decreases at high $L$, the halofit SNR is nearly constant and larger than the linear one by about an order of magnitude at $L>1000$. For low redshifts, $z\simeq 0.5$, non-linearities enter the SNR already at $L\sim 100$. This is not quite unexpected, as at least for the quadratic term, the noise is an integral over all values of $\ell$.

\renewcommand{\arraystretch}{1.5}
\begin{table}[!ht]
\centering
\begin{tabular}{|c|c||c|c|}
\hline
z   & $\Delta z$ & SNR (shot noise fixed) & SNR (shot noise adjusted)\\
\hline
1.91 & 0.05 & 11 & 8\\ 
\hline
1.91 & 0.1 & 13 & 11\\
\hline
1.91 & 0.2 & 17 & 16\\
\hline
1.91 & 0.5 & 22 & 22\\
\hline
\end{tabular}
\caption{Total SNR  for $L_{\max}=1500$ and $z=1.91$ as a function of redshift bin width for the non-linear power spectrum, assuming either a fixed number of galaxies inside all redshift bins as that given by the wide photometric bin of Euclid-like with $\Delta z = 0.5$ or adjusting the galaxy number as a function of bin width. \label{t:SNRdz}}
\end{table}
\renewcommand{\arraystretch}{1}

\begin{figure}[!ht]
    \centering
    \includegraphics[width=0.7\textwidth]{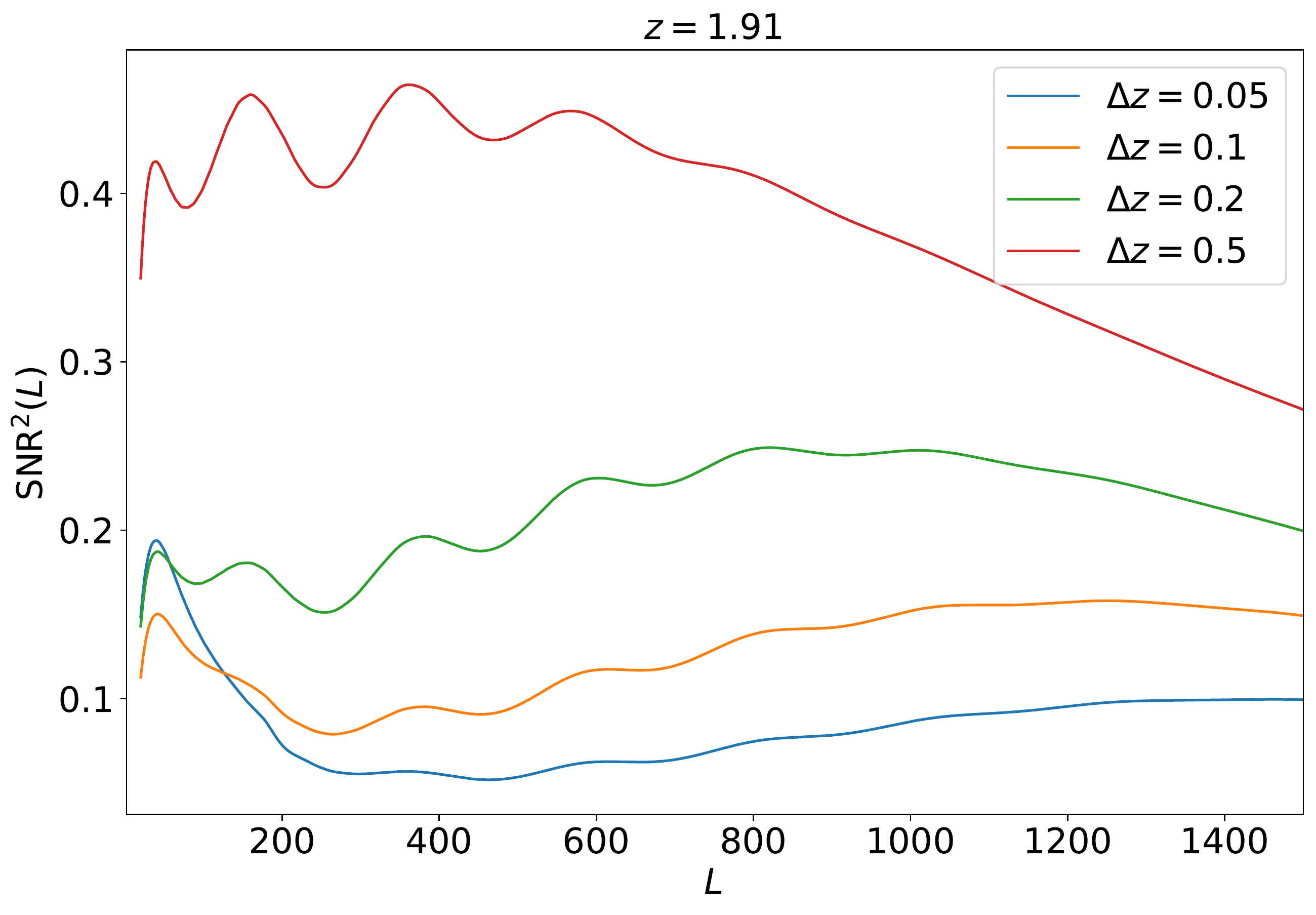}
    \caption{SNR dependence on bin width (for a fixed mean redshift $z=1.91$) for non-linear power spectra of galaxy number counts (corresponding to the 3$^{\rm{rd}}$ column of Table \ref{t:SNRdz} where the galaxy number density is not adjusted according to the bin-width).}
    \label{fig:SNR_z_1.91_function_of_binwidth}
\end{figure}

We have also studied how the SNR depends on the bin thickness. As is well known, for slimmer bins, $C_L^\De$ is larger as it is less averaged over radial modes~\cite{DiDio:2013bqa}. Since for lensing estimated from number counts the noise is dominated by the linear contribution and $N_\De^{\rm linear}= C_L/g_\De^2$ is proportional to the number count signal, we expect a smaller SNR for smaller bin width.
This is exactly what we find in Table~\ref{t:SNRdz}.
 There we determine the SNR for a fictitious survey which has a number density of $dn(z)/d\Om =0.25/$(arcmin)$^2$ at $z=1.91$ bin packed into redshift  bins of different widths $\De z$. For example, we see that when decreasing the bin width by a factor of 10, the SNR decreases by nearly a factor of 2.
If we consider a more realistic situation and also reduce the number density of galaxies proportional to the bin width, the decrease becomes nearly a factor of $3$. Therefore, relatively wider bins, where the lensing signal is a significant fraction of the total signal, are better to measure the lensing potential.
The $L$-dependence of the SNR for different bin widths at constant angular density is shown in Fig.~\ref{fig:SNR_z_1.91_function_of_binwidth}. Clearly, wider bins result in a higher SNR on nearly all scales. Narrower bins would of course allow to measure the lensing potential in more redshift bins, somewhat balancing the decreasing signal per bin.

\section{Conclusion}\label{s:con}
We have derived a new linear$+$quadratic estimator for the lensing potential from galaxy number count observations. Contrary to the CMB and intensity mapping, lensing contributes to number counts already at first order in perturbation theory. 
It turns out that this contribution greatly increases the achievable SNR for galaxy number counts compared to intensity mapping. 
We have also found, see Appendix~\ref{A:like}, that the SNR from cross-correlations typically is larger than the contribution from auto-correlations. Of course there are $n_{\rm bin}(n_{\rm bin}-1)/2$ cross-correlations and only $n_{\rm bin}$ auto-correlations for $n_{\rm bin}$ bins, but also the lensing signal is well known to be significantly more relevant in cross-correlations than in auto correlations, see~\cite{Montanari:2015rga}.

We have in particular predicted the SNR for number counts as expected from near-future photometric surveys. While within linear perturbation theory, the SNR rapidly decays for $L>200$, including non-linearities modelled by halofit keeps the SNR nearly constant in the range $400<L<1500$. For the highest redshift bins in a survey like Euclid or LSST with reduced $m_{\rm lim}$, the SNR is nearly of order unity for $L\gtrsim 100$ leading to a cumulative SNR of about 38. 
The SNR typically increases with redshift and with bin width, as for a larger redshift, the lensing signal increases, and for a larger bin width, the lensing contributes a larger part to the total signal. However, the number count SNR for lensing strongly depends on the pre-factor $2-5s(z)$ of the linear piece and is reduced to roughly the intensity mapping signal when $s(z)=2/5$. Therefore an accurate determination of the survey-specific quantity $s(z)$, defined in Eq.~\eqref{e:szF}, is crucial for the approach proposed here. In this work we have not included an uncertainty in $s(z)$ in our prediction, this is left for future work. Furthermore, as maximizing $2-5s(z)$ is crucial for a high SNR, it may be more optimal in some cases to consider a higher flux limit $F_*$ in order to increase this pre-factor, even though increasing $F_*$  reduces the number density of galaxies and therefore increases the shot noise.
This is exactly what we find when comparing LSST-like surveys with $m_{\rm lim}=25$ and $m_{\rm lim}=27$. While the latter contains more galaxies, the factor $2-5s(z)$ is significantly smaller and the total SNR for the latter is only about 5 while it is nearly 38 for the former. 

For a number density of $dn/d\Om \simeq 0.25$, we find that shot noise starts to affect the SNR at roughly $L_{\max}=1500$. Since the cosmic variance  noise (from the number counts) roughly scales as $L^{-4}$ while $C_L^{\rm SN}$ is constant, see  Fig.~\ref{fig:Noise_z_1.91_sw_0.25_linear_galaxy_IM} (where all quantities are multiplied by $L^4$), reducing the number density by  an order of magnitude will lead to shot noise starting to become significant at about $L_{\max}\simeq 1500/10^{1/4} \simeq 850$. Increasing the flux limit to optimize $s(z)$ is thus a viable strategy as long as the shot noise does not become too dominant. 

Of course one can also increase the width of the redshift bin in order to enhance to angular number density. This has the additional advantage of reducing the number count signal and thereby enhancing the SNR of lensing. However, this means that we can map the lensing potential in less redshift bins and have a more `crude' tomography.

With these caveats in mind, we are convinced that number counts provide promising direction to measure the lensing power spectrum tomographically for a wide range of $L$ values. Especially the fact that they are entirely independent of intrinsic alignment makes them a very welcome complement to shear measurements. While we have verified with a simple example that in particular the dominant linear part of our estimator works in practice, applications to more realistic simulations will be necessary to understand our approach better. We plan to study this as a next step.

\section*{Acknowledgements}
We thank Francesca Lepori for helping us with the LSST magnification bias and shotnoise.
This work is supported financially by the Swiss National Science Foundation. Mona Jalilvand acknowledges support through a McGill University postdoctoral fellowship. The numerical calculations were performed on the {\em Baobab} cluster of the University of Geneva. 

\vspace{2cm}

\noindent{\bf\Large Appendix}

\appendix
\section{Basics on quadratic estimators}\label{A:basic}
In this Appendix we derive Eqs.~\eqref{e:estX} to \eqref{e:NtotX}. We suppress the redshift dependence because the analysis is independent of it. It can be performed in each redshift bin. Let us first calculate the expectation value of $\phi_X(\bL)$ for fixed $\phi$ but stochastic $X$. We assume $\bL\neq 0$ and set $L=|\bL|$.
According to \eqref{e:estX}
\bea
\phi_X(\bL) &=& A(L)N_X(L)\int  \frac{d^2\ell}{2\pi}X(\bell)X(\bL-\bell) F_X(\bell,\bL-\bell) ~+~
(1-A(L))\frac{X(\bL)}{g_X(L)} 
\eea
where $f_X$ is defined in \eqref{e:def-fX}.
Making use of \eqref{e:Xiso} and \eqref{e:Xlens} we obtain for $\bL\neq 0$
\be\label{e:brackphi}
\lan X(\bell_1)X(\bL-\bell_1)\rangle_\phi = \frac{1}{2\pi} f_X(\bell_1,\bL-\bell_1)\phi(\bL) \,.
\ee
The factors $A$ and $(1-A)$ indicate that we consider a weighted combination of the quadratic and the linear estimator of $\phi$.
With the definitions of $F_X$ and $N_X$ we find 
\bea
\langle\phi_X(\bL)\rangle_\phi &=& \phi(\bL) \,. \label{ea:estX}
\eea

To estimate the noise we have to determine the variance  $V_X(\bL,\bL')=\langle\phi_X(\bL)\phi_X(\bL')\rangle$, now performing an ensemble average first over realizations of the variable $X$ for fixed lensing potential and after over realizations of the lensing potential. This simplifies the calculation since for the first averaging process, $\langle\cdots\rangle_\phi$ the variable $X$ can be considered as Gaussian, hence there are no terms which mix the quadratic and the linear contribution. In the second averaging procedure $\langle\phi_X(\bL)\rangle = 0$.  
\bea
V_X(\bL,\bL') &=& A(L)N_X(L)A(L')N_X(L')\times \nonumber\\
&& \qquad\int  \frac{d^2\ell d^2\ell'}{(2\pi)^2} \langle\langle X(\bell)X(\bL-\bell)X(\bell')X(\bL'-\bell')\rangle_\phi\rangle F_X(\bell,\bL-\bell)F_X(\bell',\bL'-\bell') 
\nopagebreak \nonumber\\ 
 && + ~ \frac{(1-A(L))(1-A(L'))}{g_X(L)g_X(L')}\langle\langle X(\bL)X(\bL')\rangle_\phi\rangle \,.
\eea
For $\langle\cdots\rangle_\phi$ we use \eqref{e:brackphi} and Wick's theorem, assuming $X$ to be a Gaussian variable for fixed $\phi$. As expectation values of a product of three Gaussian variables vanish, there are no mixed terms from the first and second expression. We also assume $L\neq 0$ and $L'\neq 0$ so that 
\bea
&&\langle X(\bell)X(\bL-\bell)X(\bell')X(\bL'-\bell')\rangle_\phi = ~~ \cdots\times \phi ~ + \nonumber \\
&&  \qquad\qquad  \Bigg[\frac{1}{(2\pi)^2}f_X(\bell,\bL-\bell)f_X(\bell',\bL'-\bell')\phi(\bL)\phi(\bL') +  \tilde C_\ell\tilde C_{|\bL-\bell|}\de(\bell+\bell')\de(\bL+\bL') \nonumber \\
&& \hspace*{3cm} +
\tilde C_\ell \tilde C_{|\bL-\bell|}\de(\bell+\bL'-\bell')\de(\bL-\bell+\bell') \Bigg] \,. \label{e:A4}
\eea
In the first line we schematically indicate the terms linear in $\phi$ which then drop in the expectation value over $\phi$.
Here $\tilde C_\ell$ denotes the $X$ power spectrum for fixed lensing potential $\phi$. We now take also the ensemble average over $\phi$. The first term then just becomes $\de(\bL+\bL')C_L^{\phi\phi}$ as the integrals over $\ell$ and $\ell'$ both contribute one factor of $1/N_X(L)$. We neglect this term below as it is quadratic in $\phi$ and we have neglected other terms quadratic in $\phi$. We also note that
inside the $\ell'$--integral we can substitute $\bL'-\bell'\ra\bell''$ in the third term. With this both terms contribute the same and with our definition of $F_X$ we obtain
\bea
V_X(\bL,\bL')
 &=&\de(\bL+\bL')\left\{A^2(L)\left[C_L^{\phi\phi}+N_X(L)^2\int  d^2\ell \frac{\left(f_X(\bell,\bL-\bell)\right)^2}{2C_\ell C_{|\bL-\bell|}}\right]
   ~ + ~ \frac{(1-A(L))^2}{g_X(L)^2}C_L \right\}  \nonumber \\
 & =& \de(\bL+\bL')\left(A^2(L)N_X(L) +\frac{(1-A(L))^2}{g_X(L)^2}C_L \right)  = \de(\bL+\bL') v_X(L)\,. \label{e:varX}
\eea
In the second equal sign we have neglected the quadratic term $C_L^{\phi\phi}$.
We have also inserted the expectation value of the second line of \eqref{e:A4} which gives simply $\de(\bL+\bL')C_L$.
For the last equal sign we used expression \eqref{e:noiseX} for $N_X(L)$. Note also that in Eq.~\eqref{e:varX} the $C_\ell$'s denote the true observed galaxy number count power spectrum including lensing and shot noise. To find the best combination of the linear and quadratic terms, we now choose $A(L)$ to minimize the variance $v_X$. Solving $\dd v_X/\dd A=0$ we find
\be
A(L) = \frac{C_L/g^2_X(L)}{C_L/g^2_X(L)+N_X(L)} \,.
\ee
 Inserting this expression for $A(L)$ in the variance, we find
\bea \label{eq:variancefinal}
v_X(L) &=& \frac{C_L}{\left[(C_L+g_X^2(L)N_X\right]^2}\left[C_LN_X +g_X^2N_X^2\right] \nonumber \\
\label{ea:Ntot}
&=&  \frac{C_LN_X}{C_L+g_X^2(L)N_X} \equiv N_X^{\rm (tot)}(L)\,. 
\eea
This variance is the total `reconstruction noise', $N_X^{\rm (tot)}(L)$, of the lensing power spectrum determined by this method. 
To include cosmic variance of the lensing potential, one simply has to replace $N^{\rm (tot)}_X(L)$ by $N^{\rm (tot)}_X(L)+C^{\phi\phi}_L$ in Eq.~\eqref{ea:Ntot}.

\section{Likelihood-based derivation}\label{A:like}
\newcommand{\Cov}[0]{\xi_\phi}
\begin{figure}[h!]
    \centering
    \includegraphics[width=0.7\textwidth]{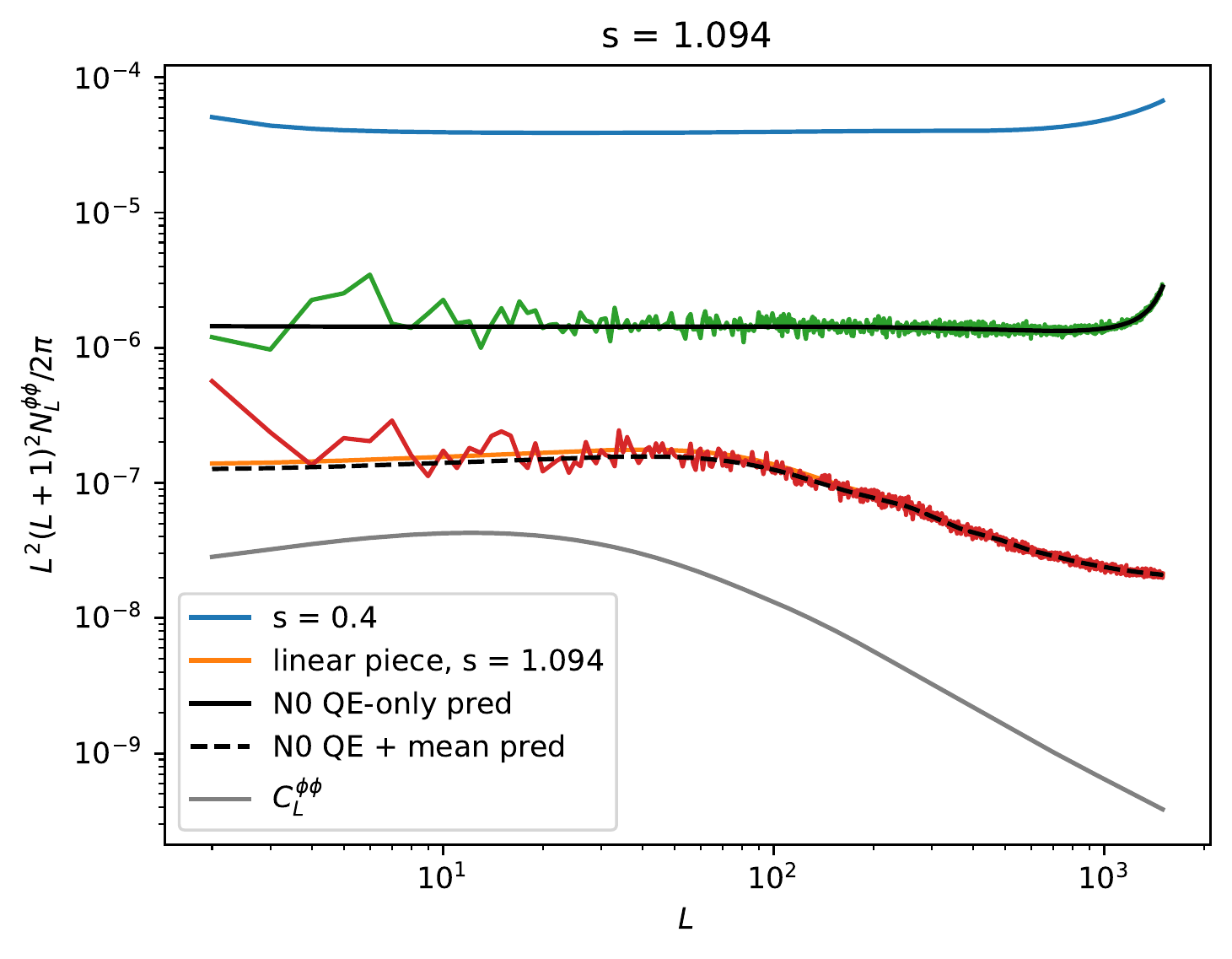}
    \caption{ Results of the application of the estimators to a toy simulation of galaxy counts (wiggly lines) at z = 1.91. A Gaussian density field is first generated from a galaxy power spectrum calculated from CAMB. Galaxy counts are then drawn following the underlying density, within pixels of a NSIDE=1024 healpy map. The pixel size is 3.2 arcmin resulting in a mean occupation number of about 3 galaxies per pixel. The linear and quadratic estimators (using $\ell_{\rm max} = 1500$) are then applied  using \texttt{plancklens} to this map of counts, which contains no signal, and shown are their power spectra. The black dashed and solid line show the reconstruction noise predictions which matches very well the recovered spectra. The blue curve shows the special case $s=0.4$, when only the lensing remapping is relevant.}
    \label{fig:QEgalsim}
\end{figure}

\begin{figure}[h!]
    \centering
    \includegraphics[width=0.8\textwidth]{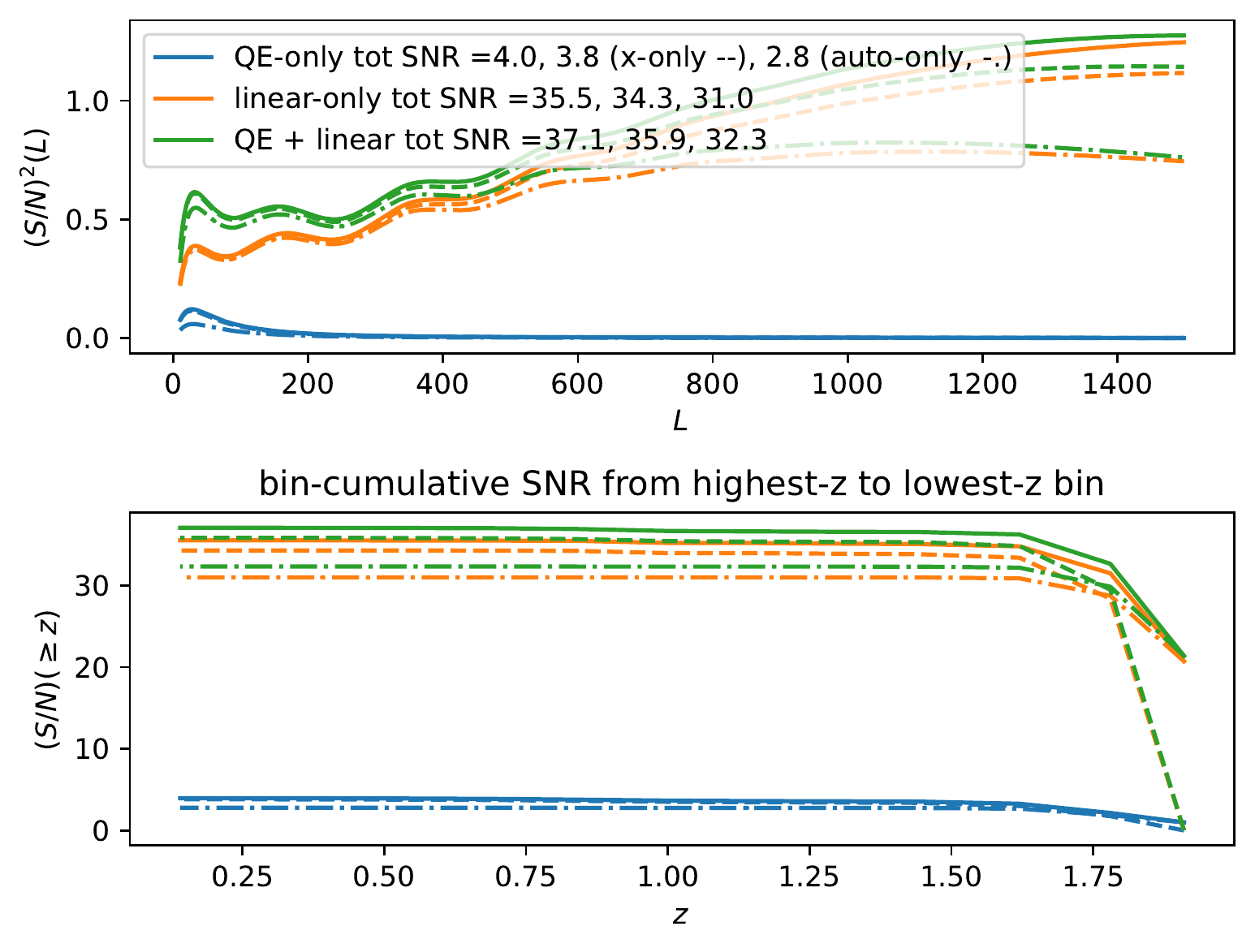}
    \caption{The SNR of the joint auto-spectra and cross-spectra using the bins presented in Table 1 for the Euclid-like number densities and magnification bias. The upper panel shows the squared SNR as function of $L$, including all the spectra and cross-spectra. The Quadratic SNR is peaked at low $L$ while the linear part peaks at high-$L$. The lower panel shows the cumulative signal to noise as function of redshift. There is barely any contribution anymore below $z\leq 1.5$. Solid lines show the total SNR, dashed including only cross-spectra, and dot-dashed only the auto-spectra.
    }
    \label{fig:EUCLIDSNRs}
\end{figure}

In this appendix we provide another derivation of the noise curves of the quadratic and linear estimators, based on a likelihood-form for the galaxy number count data. Our starting point is formula~\eqref{galaxy2ndOrder}. To the same order, and introducing $f_L = -(2-5s)\kappa_L \equiv g_L \phi_L$, we can write the density in real space as
\begin{equation}
\Delta_{g} = f(z, \bn) + 
  \tilde\Delta_{g}(z,\bn + \nabla \phi) + \tilde \Delta_{g}(z, \bn)f(z, \bn)
 \end{equation}
For fixed lenses, the overdensity field remains Gaussian, with an anisotropic covariance. The second term on the right-hand side is standard lensing remapping. The last term shares similarities to the case of a modulation field as searched for instance in the CMB.  Additionally, and different to CMB lensing, the mean of the Gaussian overdensity field is non-zero (for fixed lenses, ensemble averaging we have $\left\langle \Delta_g  \right\rangle = f$ the first term). Under this Gaussian model, the optimal estimator for $\phi$ will be built out two pieces, a quadratic estimator probing the anisotropic covariance, and a linear piece probing the mean. The two estimators will be independent to leading order, so that the resulting total noise is given by inverse variance weighting of the QE noise ($N_L$, or $N_L^{(0)}$) and that of the linear piece ($C_L / g_L^2$) . The quadratic piece itself contains two elements: the `standard' lensing deflection anisotropy estimate, and that of the modulation field estimate. For small anisotropies, we can derive all these quantities using a likelihood-based approach (see e.g. \cite{Hanson:2009gu} for a detailed derivation). The lensing signal likelihood is
\begin{equation}
    \ln p (\Delta_g | \phi) = -\frac 12 \left(\Delta_g - g\phi \right) \Cov^{-1} \left(\Delta_g - g\phi \right) - \frac 12 \ln |\Cov|
 \end{equation} where $\Cov$ is the anisotropic number count spectrum (or anisotropic two-point function in a position-space description), reducing to $C_\ell$ for vanishing $\phi$. Then, the (unnormalized) optimal estimator is given by the gradient $ -\delta \ln p / \delta \phi^\dagger_L $ and its normalization (the response $\mathcal R_L$, equal to the Fisher information matrix $F_L$, itself equal to the inverse Gaussian reconstruction noise level $N_L$), is its second variation $\left\langle - \delta^2 \ln p / \delta \phi_L \delta \phi^\dagger_L \right \rangle$. These quantities are evaluated for vanishing anisotropies, resulting in isotropic weights and noise levels. In practice, for CMB lensing, the signal contributes to the noise levels and a more accurate estimate including non-perturbative effects is given by replacing the unlensed by lensed spectra~\cite{Hanson:2010rp}. Here the impact of the lensing on the power spectra is smaller. 
The normalized estimator is thus
\begin{equation}
    \hat \phi_{\boldsymbol{L}} = \frac{1}{\mathcal R_L} \left(\frac{g \Delta_{g,\boldsymbol{L}}}{C_L} - \left(\frac{ \Delta_g}{C} \right)_{\bell_1}\left(\frac{\delta \Cov}{\delta \phi^\dagger_L}\right)_{\bell_1\bell_2,\phi = 0} \left(\frac{ \Delta_g}{C} \right)_{\bell_2}\right),
\end{equation}
where the first part comes from the variation of the mean, and the second from the inverse covariance. The determinant part gives no contribution by symmetry, unless non-idealities such as masking or anisotropic noise are taken into account, in which case it is usually called the `mean-field'.
The quadratic piece is the sum of the standard lensing QE for the remapping part of the signal and unnormalized modulation field QE for `f'. The full response is
\begin{equation}
\mathcal R_L  =  \mathcal R_L^{\phi, \phi} + g_L \mathcal R_L^{\phi, f} + g_L \mathcal R_L^{f, \phi} + g_L^2 \mathcal R_L^{f, f}  + g_L^2\frac{1}{C_L} \equiv \frac{1}{N_L}+ g_L^2\frac{1}{C_L}
\end{equation}
where $\mathcal R^{a, b}_L $ stands for the response of QE $a$ to anisotropy source $b$ and is given by $\left \langle {-\delta^2 \ln p / \delta a_L \delta b^\dagger_L} \right \rangle $. The total noise is given by standard inverse-variance weighting,
\begin{equation}
    N_L^{\rm tot} = \mathcal R_L^{-1} = \left( \frac{1}{N_L} + \frac{g_L^2}{C_L} \right)^{-1} \,,
\end{equation}
in agreement with \eqref{ea:Ntot}.

The \texttt{plancklens} package~\url{www.github.com/carronj/plancklens} contains all the necessary calculations of the responses implemented here. This package uses the full curved-sky formalism and is not limited to $L>20$. With the help of this package we have also created a modulated and Poisson-sampled example galaxy field and we have verified that we could recover the corresponding convergence spectrum with our estimator, see Fig.~\ref{fig:QEgalsim}. We also show the cummulative SNR starting at the highest redshift in Fig.~\ref{fig:EUCLIDSNRs}. Note that while both, auto- and cross-spectra contribute with the same order of magnitude, the SNR from the cross-spectra is higher. Already in~\cite{Montanari:2015rga} is was noted, that cross-spectra are especially sensitive to the lensing signal.

\section{Thermal noise}\label{C:noise}
In this appendix we reproduce the expression for thermal noise of intensity mapping.
In Eq.\ (D2) of ~\cite{Bull_HI_noise}, the thermal noise is given in Fourier space ($\boldsymbol{\ell} = \mathbf{k}_{\perp} r,y=k_{\|} c(1+z)^2/H(z)$) for a redshift bin centered at $z$ as 
\be  
C^{N}({\ell}, y)= \left (\frac{T_{\rm sys}(z)}{\bar{T}(z)} \right) ^2   \frac{\lambda (z)^{4} S_{\text {area }}}{\nu_{21} n_{\mathrm{pol}} t_{\text {tot }} A_{\mathrm{eff}}^{2} \cdot \mathrm{FOV} N_{\mathrm{beam}} n( \ell ) }  \, ,
\ee
where $T_{\mathrm{sys}}=T_{\mathrm{antenna}}+T_{\mathrm{sky} }$, $\bar{T}(z)$ is the mean 21cm temperature at redshift $z$, $\nu_{21}$ is the 21cm radiation rest frame frequency, $\mathrm{FOV} = \theta_b^2$ where $\theta_b = \lambda(z) / D_{\mathrm {dish} }$ is the beam of the telescope, $S_{\mathrm{area}}= 4\pi f_{\mathrm{sky} }$, $\lambda(z) $ is the observed wavelength of 21cm radiation at
redshift $z$, $n_{\mathrm {pol}}$ is the number of polarizations, $t_{\mathrm {tot}}$ is the total observation time, $N_{\mathrm{ beam} }$ is the number of beams, $A_{\mathrm{eff} } = 0.7 \pi D_{\mathrm{dish} }^2/4 \,$ is the effective area of each dish and the factor $0.7$ is the efficiency of the dish, and $n(\ell)$ is the baseline number density in $\ell$-space. Denoting the frequency bin corresponding to the redshift bin by $\Delta \nu$, the relation between the two dimensional power-spectrum  $C_\ell$ and $C({\ell},y)$ is given by
\be C(\ell)=\frac{C({\ell}, y) \nu_{21}}{\Delta \nu} \, ,\ee
and therefore the $2\mathrm{D} $ noise spectrum is
\be
C^{\rm interf}_\ell
(z) = \left (\frac{T_{\rm sys}(z)}{\bar{T}(z)} \right) ^2 \frac{ S_{\rm area} \, \lambda(z)^4}{n_{\rm pol} \, t_{\rm tot} \, \Delta \nu \, N_{\rm beam} \, A_{\rm eff}^2 \, \theta _b^2 \, n(\ell)}\, .
\label{eq:noiseBULL}
\ee

\renewcommand{\arraystretch}{1.2}
\begin{table}[h!]
\centering
\begin{tabular}{|c|c|}
\hline
Paramaters   & Values\\
\hline
$D_{\rm dish}$ & 6 [m]\\
\hline
$T_{\rm antenna}$& 50[K]\\ 
\hline
$T_{\rm sky}$ & $60 \rm K (\nu(z)/300\rm MHZ)^{-2.55}$\\ 
\hline
$n_{\rm pol}$& 2\\
\hline
$S_{\rm area}$& 15000 [$\rm deg^2$]\\
\hline
$t_{\rm tot}$& 2.8 years (optimistic)\\
\hline
$N_{\rm beam}$ & 1\\
\hline
$N_{\rm dish}$ & 1024\\
\hline
\end{tabular}
\caption{Specifications of HIRAX intensity mapping survey.\label{tab:hirax}}
\end{table}
\renewcommand{\arraystretch}{1}

For a compact square interferometric survey, the following fitting formula can be used for the baseline number density \cite{Durrer:2020orn}
\be n (z, \ell)=N_{\mathrm{d}}\left(\frac{\lambda(z)}{D_{\mathrm{dish}}}\right)^{2}\left[\frac{a_{1}+a_{2}\left(L / L_{\mathrm{s}}\right)}{1+a_{3}\left(L / L_{\mathrm{s}}\right)^{a_{4}}}\right] \exp \left[-\left(\frac{L}{L_{\mathrm{s}}}\right)^{a_{5}}\right] \, ,
\ee 
where $N_d$ is the number of dishes, $L_{\mathrm{s}}=D_{\mathrm{d}} \sqrt{N_{\mathrm{d}}}$, and $L$ is
\be L(z, \ell)=\frac{\lambda(z)}{2 \pi} \ell \, . \ee 
The numerical values of the quantities that we used to model the HIRAX IM survey are given in Table \ref{tab:hirax}.

\section{Data for galaxy number counts} 
In Table~\ref{tab:nz_all} we present the galaxy number density ($dn(z)/d\Om$), the  magnification bias ($s(z)$), and the galaxy linear bias ($b(z)$) for the photometric surveys used in this paper. The specifications for the LSST-like photometric survey are available in the \href{http://intensitymapping.physics.ox.ac.uk/Codes/ULS/photometric/}{code} used in \cite{Alonso:2015uua} where we have used all the galaxies (red and blue). For the Euclid-like survey, we use the specifications given in \cite{Euclid:2021rez}.

\renewcommand{\arraystretch}{1.2}
\begin{table}[h!]
    \centering
    \begin{tabular}{|c|c||c|c|c||c|c|c||c|c|}
    \hline
       &   &  \multicolumn{3}{c||}{$dn(z)/d\Omega$ [arcmin$^{-2}$]} & \multicolumn{3}{c||}{$s(z)$} & \multicolumn{2}{c|}{$b(z)$}   \\ \cline{3-10}
      $z$  & $\Delta z$ & Euclid & LSST & LSST & Euclid & LSST & LSST & Euclid & LSST \\
        &  &   & \small{$m_{\lim}=25$} & \small{$m_{\lim}=27$}  &  & \small{$m_{\lim}=25$}  & \small{$m_{\lim}=27$}  &  &   \\ \hline\hline
    0.14 & 0.2 & 0.76 &  1.88 &  3.68 &  0.02 &  0.15 &  0.14 &  0.62 &  1.12\\ \hline
    0.26 & 0.2 & 2.61 &  3.26 &  6.64 &  0.14 &  0.16 &  0.15 &  0.92 &  1.22\\\hline
    0.39 & 0.2 & 4.12 &  4.14 &  8.84 &  0.25 &  0.18 &  0.15 &  1.12 &  1.33\\\hline
    0.53 & 0.2 & 3.84 &  4.42 &  10.01 &  0.25 &  0.20 &  0.16 &  1.35 &  1.45\\\hline
    0.69 & 0.2 & 3.86 &  4.12 &  10.19 &  0.23 &  0.23 &  0.17 &  1.54 &  1.58\\\hline
    0.84 & 0.2 & 3.73 &  3.50 &  9.60 &  0.28 &  0.26 &  0.19 &  1.60 &  1.71\\\hline
    1.00 & 0.2 & 3.00 &  2.71 &  8.51 &  0.39 &  0.31 &  0.20 &  1.84 &  1.84\\\hline
    1.14 & 0.2 & 2.83 &  2.07 &  7.41 &  0.48 &  0.36 &  0.22 &  1.85 &  1.96\\\hline
    1.30 & 0.2 & 1.80 &  1.45 &  6.13 &  0.60 &  0.43 &  0.24 &  2.10 &  2.09\\\hline
    1.44 & 0.2 & 1.08 &  1.00 &  5.06 &  0.79 &  0.50 &  0.26 &  2.27 &  2.21\\\hline
    1.62 & 0.2 & 0.52 &  0.57 &  3.81 &  1.06 &  0.62 &  0.29 &  2.48 &  2.36\\\hline
    1.78 & 0.5 & 0.36 &  0.85 &  7.28 &  1.14 &  0.76 &  0.32 &  2.19 &  2.50\\\hline
    1.91 & 0.5 & 0.25 &  0.49 &  5.63 &  1.09 &  0.91 &  0.36 &  2.16 &  2.60\\\hline
    2.17 & 0.5 &  -   & 0.12  & 3.11  &  -    & 1.33  & 0.45  &  -   & 2.82 \\\hline
    2.43 & 0.5 &  -   & 0.02  & 1.47  &  -    & 2.01  & 0.59  &  -   & 3.04 \\\hline
    \end{tabular}
    \caption{The galaxy number density per bin ($dn(z)/d\Omega$ [arcmin$^{-2}$]), the selection magnification bias ($s(z)$), and the selection bias ($b(z)$) for all the redshift bins for the Euclid-like and LSST-like surveys that we consider. The redshift bin widths are given in the second column.}
    \label{tab:nz_all}
\end{table}
\renewcommand{\arraystretch}{1}

\bibliographystyle{JHEP}
\bibliography{ref}

\end{document}